**Spin-Orbit Coupled Exciton-Polariton Condensates in Lead Halide Perovskites**

Michael S. Spencer[1] *, Yongping Fu[1] *, Andrew P. Schlaus[1], Doyk Hwang[1], Yanan Dai[1], Matthew D. Smith[2], Daniel R. Gamelin[2], and X.-Y. Zhu[1] †

[1] Department of Chemistry, Columbia University, New York, NY 10027, USA

[2] Department of Chemistry, University of Washington, Seattle, Washington 98195-1700, USA

*These authors contributed equally to this work.
†Author to whom correspondence should be addressed: xyzhu@columbia.edu

**ABSTRACT.** Spin-orbit coupling (SOC) is responsible for a range of spintronic and topological processes in condensed matter. Here we show photonic analogs of SOCs in exciton-polaritons and their condensates in microcavities composed of birefringent lead halide perovskite single crystals. The presence of crystalline anisotropy coupled with splitting in the optical cavity of the transverse electric (TE) and transverse magnetic (TM) modes gives rise to a non-Abelian gauge field, which can be described by the Rashba-Dresselhaus Hamiltonian near the degenerate points of the two polarization modes. With increasing density, the exciton polaritons with pseudospin textures undergo phase transitions to competing condensates with orthogonal polarizations. Unlike their pure photonic counterparts, these exciton polaritons and condensates inherit nonlinearity from their excitonic components and may serve as quantum simulators of many-body SOC processes.

**INTRODUCTION**

Spin-orbit coupling (SOC) of electrons is responsible for a number of quantum phenomena in solids, including spin Hall effect(*1*) and topological insulators(*2*). Such SOC phenomena also emerge in photonic systems(*3–8*). In planar microcavities, the natural splitting between transverse electric (TE) and transverse magnetic (TM) modes behaves as a winding in-plane magnetic field $\vec{B}_T$ on the photon spin and results in the photonic spin-Hall effect(*4–6*). If in-plane optical anisotropy is present to break rotational symmetry(*9*), there is an effectively constant magnetic field $\vec{B}_{XY}$ which, in combination with $\vec{B}_T$, leads to a non-Abelian gauge field for photons or exciton-polaritons near the so-called diabolical point, where the fields cancel(*10*). There have been increasing interests in recent years to introduce spin-orbit coupling directly via engineering a photonic system to exhibit optical analogs of SOC and topological effects from pseudo-magnetic fields and/or pseudo-spins (*11–13*). While such photonic systems have been demonstrated for synthetic Rashba-Dresselhaus Hamiltonians(*3, 8, 14–16*), an exciting prospect is forming exciton-polariton condensates(*17, 18*) under such an effective field to simulate a number of phenomena in SOC quantum fluids(*10*), in analogy to what have been demonstrated in SOC Bose-Einstein condensates (BECs) in cold atoms(*19–21*). Here we demonstrate spin-polarized exciton-polaritons and condensates in microcavities composed of single crystal lead halide perovskites (LHPs)(*22*) known for strong light-matter coupling in microcavities (*23–27*) including topological photonic structures (*27, 28*). We take advantage of low-symmetry phases of LHPs with strong optical anisotropy, which is tunable by composition and/or temperature(*29*), and measure the spin textures produced by $\vec{B}_T + \vec{B}_{XY}$ using polarization resolved Fourier space photoluminescence (FS-PL) imaging. In the present study, we focus on two regions in the in-plane momentum space ($k_\parallel$): i) the region near the so-called diabolic point ($k_{db}$) where $\vec{B}_T$ and $\vec{B}_{XY}$ cancel each other and the Hamiltonian is of the Rashba-Dresselhaus form; and ii) the region near $k_\parallel = 0$ where condensates form from the spin-polarized exciton-polaritons as their density reaches a threshold. The present study will not probe propagating condensates near $k_{db}$; this will be subject of future studies.

In a planar DBR cavity with the presence of TE-TM splitting and optical anisotropy, the Hamiltonian describing the eigenstates in a circular polarization basis is given by (*7, 10*):

$$\boldsymbol{H}_{Ph} = \begin{pmatrix} E_0 + \frac{\hbar^2 k_\parallel^2}{2m} & -\alpha e^{-i\varphi_0} + \beta k_\parallel^2 e^{-2i\varphi} \\ -\alpha e^{i\varphi_0} + \beta k_\parallel^2 e^{2i\varphi} & E_0 + \frac{\hbar^2 k_\parallel^2}{2m} \end{pmatrix} \quad (1)$$

where the degenerate diagonal terms are cavity photon modes. $k_\parallel$ is in-plane momentum with propagation angle $\varphi$, $\boldsymbol{k}_\parallel = (k_\parallel \cos\varphi, k_\parallel \sin\varphi)$; $E_0$ is the mode energy at $k_\parallel = 0$ and $m$ the cavity reduced mass for a given mode; $\beta k_\parallel^2$ and $\alpha$ represent the strength of TE-TM splitting and optical anisotropy, respectively; $\varphi_0$ is the fixed in-plane angle of $\vec{B}_{XY}$ with respect to the optical axis. As a 2×2 Hermitian Hamiltonian, equation (1) can be decomposed into a linear combination of Pauli matrices, which can be viewed as Zeeman interaction between an effective magnetic field and the pseudospin of a photon(*7*) (Supplementary text 1):

$$\boldsymbol{H}_{Ph}(\boldsymbol{k}_\parallel) = \left(E_0 + \frac{\hbar^2 k_\parallel^2}{2m}\right)\mathbb{1} + \overrightarrow{\boldsymbol{G}(\boldsymbol{k}_\parallel)} \cdot \vec{\sigma} \quad (2)$$

where $\mathbb{1}$ is the 2×2 unit matrix, $\vec{\sigma}$ is a vector of Pauli Matrices; $\vec{G}(k_\parallel) = \left(-\alpha\cos(\phi_o) + \beta k_\parallel^2 \cos(2\phi), -\alpha\sin(\phi_o) + \beta k_\parallel^2 \sin(2\phi), 0\right)$ is the effective magnetic field given by the sum of $\vec{B}_{XY}$ and $\vec{B}_T$. The pseudospins of the eigenstates are either aligned or anti-aligned with the effective field. The $k_\parallel$-dependent effective field acting on the photon pseudospin can be physically interpreted as a photonic SOC around the diabolic points.

**RESULTS AND DISCUSSIONS**

We directly grow single crystal microplates of MAPbBr$_3$ (MA = methylammonium) or CsPbBr$_3$ perovskites in an empty cavity formed by two laminated distributed Bragg reflectors (DBRs), Figs. 1A and 1B. The natural thickness gradient of the microcavities allows us to tune the cavity resonances across the exciton resonance ($E_{ex}$). The as-grown single crystals exhibit 10-100µm in lateral dimensions and a few micrometers in thickness. Additional optical images (Fig. S1), powder X-ray diffraction (Fig. S2), atomic force microscopy (Fig. S3), and reflectance spectra (Fig. S4) of representative samples are provided in the Supplementary Information. At room temperature, CsPbBr$_3$ is in the birefringent orthorhombic structure whereas MAPbBr$_3$ is in the isotropic cubic structure(29), and this difference results in distinctive dispersions with pseudospin textures of the resulting polaritons.

We perform FS-PL imaging on the microcavities (Fig. S5) to obtain dispersion in parallel momentum, $k_\parallel = (k_x, k_y)$. Constant energy cross sections in the energy-momentum space are shown in Fig. 1C and 1D for MAPbBr$_3$ and CsPbBr$_3$, respectively, and in supplementary videos (Movie S1 and S2). The effective cavity lengths support multiple longitudinal Fabry–Pérot modes. The modes of cubic MAPbBr$_3$ are perfectly concentric in $k_\parallel$ (Fig. 1C). By contrast, optical anisotropy in CsPbBr$_3$ splits each mode into two ellipses that are degenerate at certain momentum points. The dispersion of the isotropic MAPbBr$_3$ microcavity is characterized by two orthogonal polarization modes degenerate at $k_\parallel = 0$, but split at higher $k_\parallel$ (Fig. S6), indicting the presence of the winding $\vec{B}_T$ only. By comparison, in-plane anisotropy in CsPbBr$_3$ leads to a clear mode splitting at $k_\parallel = 0$ due to $\vec{B}_{XY}$, Fig. 1E. As the cubic MAPbBr$_3$ undergoes phase transitions at lower temperature, the dispersion dramatically changes (Fig. S7). The tetragonal phase exhibits a small mode splitting due to a finite anisotropy, whereas the orthorhombic phase shows much larger $k_\parallel = 0$ splitting that is even larger than those in CsPbBr$_3$.

The high quality-factor (Q = 970±100, Fig. S8, derived from the polariton linewidth below condensation threshold) of our microcavities ensures sufficient coupling of the excitons to the photonic modes, forming SOC polaritons. Diagonalization of the Hamiltonian (1) gives two $\varphi$-dependent SOC photonic modes(*10*),

$$E_{ph\pm}(\mathbf{k}_\parallel) = E_0 + \frac{\hbar^2 k_\parallel^2}{2m} \pm \sqrt{\beta^2 k_\parallel^4 - 2\alpha\beta k_\parallel^2 \cos(2\varphi - \phi_o) + \alpha^2} \qquad (3).$$

When the SOC cavity photons are strongly coupled to an exciton resonance with Rabi splitting $\Omega$, and neglecting the difference in the exciton energies of the two orthogonal polarizations (Fig. S9), we obtain two spin-split lower polariton branches (LPBs) and two spin-split upper polariton branches (UPBs) which are given by

$$E_{Pol}(\mathbf{k}_\parallel) = \frac{E_{ex} + E_{ph\pm}(\mathbf{k}_\parallel)}{2} \pm \frac{1}{2}\sqrt{[E_{Ph\pm}(\mathbf{k}_\parallel) - E_{ex}]^2 + \Omega^2} \qquad (4).$$

Because the effective field depends exclusively on the momentum vector, not on the energy, the spin texture of each photon mode is preserved in the polariton mode (Fig. S10 and Supplementary text 2).

The excitonic and photonic fractions depend on the cavity detuning $\delta = E_0 - E_{ex}$. When $|\delta| > \Omega$, the polariton modes are more photon-like for the momentum space in our imaging range, and the energy dispersion can be described by equation (3), as shown by dashed curves on the lower energy modes in Fig. 1E. For these modes, the polariton emission strongly increases toward $E_{ex}$ at high $k$ due to the relaxation bottleneck.(*30*) Along $k_y$, Fig. 1E (left) the two anisotropic modes cross at $k_y = \pm(\alpha/\beta)^{0.5} = \pm 3.3$ μm$^{-1}$, i.e., the diabolical points ($k_{db}$) where $\vec{B}_{XY}$ and $\vec{B}_T$ cancel each other. Along $k_x$, Fig. 1E (right), the two fields add up, and the energy splitting increases with $k_x$. We reproduce the two dispersions using equation (3) with $E_0 = 2132.4 \pm 0.6$ meV, $\frac{\hbar^2}{2m} = 2.24 \pm 0.08$ meV μm$^{-2}$, $\beta = 0.34 \pm 0.03$ meV μm$^{-2}$, and $\alpha = 7.4 \pm 0.8$ meV. The two modes are orthogonally polarized (Figs. S11-S12), and the pseudospins switch across the diabolical points.

As shown in Fig. 1F, the addition of $\vec{B}_{XY}$ (green arrows) to $\vec{B}_T$ (orange arrows) breaks the rotational symmetry of the TE-TM doublets(*31*), separating the TE-TM of $4\pi$ winding into a pair of 2D monopoles of $2\pi$ winding. The effective magnetic field vanishes at the angles corresponding to $\varphi = \phi_o/2$, and at wavevectors $k_{db} = \sqrt{\alpha/\beta}\left(\cos\frac{\phi_o}{2}, \sin\frac{\phi_o}{2}\right)^T$, giving rise to the diabolical points. In the region near these points, the Hamiltonian (1) can be rewritten (Supplementary text 3) as a Rashba-Dresselhaus Hamiltonian(*10*):

$$\boldsymbol{H}_{Ph}(\boldsymbol{k}_\| = \boldsymbol{k}_{db} + \boldsymbol{q}) \approx \left(E_o + \frac{\hbar^2(\boldsymbol{k}_{db}+\boldsymbol{q})^2}{2m}\right)\mathbb{1} + \beta q^2 + 2\sqrt{\alpha\beta}\left[\cos\left(\frac{\phi_o}{2}\right)\boldsymbol{\sigma}\cdot\boldsymbol{q} + \sin\left(\frac{\phi_o}{2}\right)\boldsymbol{\sigma}\times\boldsymbol{q}\right] \quad (5).$$

where $\boldsymbol{q} = \boldsymbol{k}_\| - \boldsymbol{k}_{db}$. Note that it is within this context that the effective magnetic field produces a non-Abelian Gauge Field, whereas in general it is an effective magnetic field. For our spin-split polariton modes where the diabolic points are within the measurable $k_\|$ range, we compare the experimental pseudospin textures with predictions from the Hamiltonian (1). We map the Stokes vectors of the two polariton eigenstates and plot their $k_\|$ dependent phases $\phi_{stokes} = \tan^{-1}(S_2/S_1)$, where $S_1$ and $S_2$ are the Stokes parameters on the Poincaré sphere (Supplementary text 4). We also note that the dispersions near the diabolic point resembles the behavior near the Dirac point in graphene, with an additional slope. Indeed, the Hamiltonian in (5) is similar to the tight binding Hamiltonian of graphene, except with an additional 'twist' which reflects the fact that the gauge field can be written as a divergence, a curl, or some combination of the two, depending on the crystal angle ($\phi_o$) (see Supplemental text 3).

We show constant energy cuts of PL intensity (Figs. 1$G_1$, 1$H_1$, 1$I_1$), phase ($G_2$, $H_2$, $I_2$), and theoretical phase ($G_3$, $H_3$, $I_3$) at energies below (G), at (H) and above (I) the diabolical points, where $|\vec{B}_{XY}| > |\vec{B}_T|$, $\vec{B}_{XY} = -\vec{B}_T$, and $|\vec{B}_{XY}| < |\vec{B}_T|$, respectively. In all three regions, the experimentally retrieved pseudospin textures agree with the theoretical predictions from Hamiltonian (1). In the XY-dominant regime, a nearly constant magnetic field generates Zeeman splitting, giving rise to two modes polarized along $\phi \sim \phi_o$ and $\sim \phi_o + \frac{\pi}{2}$, respectively. In the TE-TM dominant regime, the two modes show the expected $4\pi$ phase winding. At the energy of the diabolical points, exchanging the interior and exterior degenerate eigenstates across the diabolical points gives rise to two circular rings with exactly $2\pi$ phase windings. One can observe the pseudospin points to opposite directions at each side of the points, which agrees with the Rashba-Dresselhaus field. Moreover, the Rashba energy structure is confirmed in Fig. 1J, which displays the measured dispersion intersecting the diabolical point in the direction perpendicular to $\boldsymbol{k}_{db}$. The two spin-split parabolas can be well described by the theoretically predicted $E_\pm \approx E(k_{db}) + \frac{\hbar^2 q^2}{2m} \pm 2\sqrt{\alpha\beta}q$ (the solid lines in Fig. 1J). The anisotropic splitting from $\vec{B}_{XY}$ in MAPbBr$_3$ in the low temperature orthorhombic phase (Fig. S13) is even stronger than that in CsPbBr$_3$; both are two orders of magnitude higher than those in GaAs and CdTe microcavities(*7*, *32*, *33*). The magnitudes of Rashba-Dresselhaus splitting of the exciton-polaritons in our lead halide perovskite microcavities are at levels previously only seen for pure photonic modes(*3*, *8*). Unlike the reported SOC photonic modes that possess no matter components and therefore no nonlinearity(*3*, *8*), the SOC exciton-polaritons in our lead halide perovskite microcavities are highly nonlinear and undergo phase transitions to competing condensates, as we establish below.

For polaritons with $E_0$ close to $E_{ex}$, we observe deviations from parabolic dispersions at high $k$ and avoided crossings with the exciton resonance, as seen for both MAPbBr$_3$ (Fig. S6) and the CsPbBr$_3$ microcavities (Fig. 1D and Fig. S14 at different δ values). The dispersions can be reproduced by the LPBs in equation (4) with $E_0$ = 2296.4 ± 0.6 meV, $\frac{\hbar^2}{2m}$ = 1.08 ± 0.05 meV μm$^{-2}$, $\beta$ = 0.10 ± 0.01 meV μm$^{-2}$, $\alpha$ = 10 ± 1 meV, and Ω = 12 ± 2 meV, giving rise to diabolical points at $k_y$ ~ 7.1 μm$^{-1}$. The two polariton modes are orthogonally polarized (Figs. S11-S12), in agreement with a report by Bao et al. on a CsPbBr$_3$ microcavity(*25*). Note that these authors did not observe diabolic points, spin textures, or phase transition into competing condensates.

Polariton formation becomes more obvious at lower temperatures, shown here by dispersions from a CsPbBr$_3$ microcavity at $T$ = 77, 120, 200, and 290 K, respectively, Fig. 2A$_{1-4}$ and Fig.S15. The dispersions of the LPBs closest to $E_{ex}$ are flattened at large $k_\parallel$ due to avoided crossing. Our modeling with equation (4) yields Ω = 25 ± 3, 25 ± 3, 18 ± 2, 16 ± 2 meV for this cavity at 77, 120, 180, and 290 K, respectively. The strong coupling is further supported by a clear cavity detuning effect seen in Fig. S16, which shows a series of dispersions for positive, resonant, and negative detuning. The polariton modes inherit the spin textures from the cavity photon modes, as shown by the dispersions at 77 K of horizonal (Fig. 2B$_1$), vertical (Fig. 2B$_2$), and 45° (Fig. 2B$_3$) polarizations. Polarization angular dependences of the two polariton modes near E$_{ex}$, Fig. 2B$_4$, confirm that the two spin-split polariton modes possess orthogonal polarizations.

The coupling constant Ω obtained here is more than a factor of two smaller than the Ω = 60 meV reported by Su et al. for a room temperature CsPbBr$_3$ microcavity(*24*). In addition to differences in cavity structure and crystal growth procedures between our work and that of Su et al, the different Ω values may likely come from the different $E_{ex}$ values used. We use $E_{ex}$ = 2.353±0.005 eV in our analysis. By examining the dispersions at different cavity detuning in our wedged samples, we consistently observe the avoided crossing at $E_{ex}$ = 2.353 ± 0.005 eV. Note that, unlike the nearly constant PL peak energy, the absorption peak energy blue shifts with increasing temperature (*34*), as is confirmed in our reflectance spectra (Fig. S4). This blue shift is also accompanied by significant peak broadening with temperature, giving rise to large uncertainty in the determination of $E_{ex}$ at room temperature. If we use the same $E_{ex}$ = 2.407 eV as in reference (*24*), our fitting yields $\Omega$ = 50 ± 5 meV, in closer agreement with Su et al. (*24*).

Having established the SOC exciton polariton modes with spin textures, we now turn to their phase transitions into competing condensates. Fig. 3A, 3B, 3C show PL spectra at 77 K as functions of pump fluence (*P*) and the calculated exciton density ($n_{ex}$, Supplementary text 4) for CsPbBr$_3$ microcavities with increasing detuning $\delta = E_0 - E_{ex}$ = +1 meV, −15 meV, and −40 meV, respectively. For the most resonant cavity, $\delta$ = 1 meV (Fig. 3A), we observe two-thresholds in the appearance of sharp PL peaks assigned to lasing, at $P_{th1}$ = 5.0 ± 0.3 μJ cm$^{-2}$ and $P_{th2}$ = 53 ± 3 μJ cm$^{-2}$, corresponding to $n_{ex}$ ~1×10$^{17}$ cm$^{-3}$ and ~1×10$^{18}$ cm$^{-3}$, respectively. With increased detuning, $\delta$ = −15 meV (Fig. 3B), the two lasing thresholds up-shift to $P_{th1}$ = 8.6 ± 0.4 μJ cm$^{-2}$ and $P_{th2}$ = 84 ± 4 μJ cm$^{-2}$. For the largest detuning, $\delta$ = −40 meV (Fig. 3C), there is mainly one threshold which is slightly different for the two SOC modes, $P_{th1\_a}$ = 10.6 ± 0.5 μJ cm$^{-2}$ and $P_{th1\_b}$ = 13.2 ± 0.7 μJ cm$^{-2}$ at ~537 and ~540 nm, respectively. The insets in Fig. 3A, 3B, 3C show representative spectra, black, blue, and red (or green), for $P < P_{th1}$, $P_{th1} < P < P_{th2}$ (or $P_{th1\_a} < P < P_{th1\_b}$), and $P > P_{th2}$ (or $P > P_{th1\_b}$), respectively. For $P < P_{th1}$, the broad spectrum (525-550 nm) is spontaneous emission, and the linewidth of the polariton is 0.55 nm. For $\delta$ = +1 or −15 meV, a sharp lasing peak in one or both SOC modes at ~533 nm appears with a two-orders of magnitude decrease in full width at half-maximum (FWHM) to 0.21 ± 0.01 nm, corresponding to an effective quality factor of Q = 2500. When $P > P_{th2}$, two additional lasing peaks emerge at the lower energy cavity modes (~547 nm). For the largest detuned cavity, $\delta$ = −40 meV, the second threshold for the red-shifted lasing modes is not observed within the fluence range. Instead, we observe the lower SOC mode appearing slightly delayed and increasing after the higher SOC mode has saturated.

A two-threshold lasing behavior has been considered as strong evidence for exciton-polariton condensation: the first is attributed to stimulated scattering to form the condensates, also called *polariton lasing*(*17*, *18*), and the second, at density above the Mott threshold, to stimulated emission from electron-hole plasmas (*35–39*). Polariton lasing requires the system to remain in strong coupling. With increasing $n_{ex}$, many-body screening reduces the exciton binding energy, causing the system to undergo Mott transition to an electron-hole (e/h) plasma with a reduction of the oscillator strength and resulting in weak coupling to photons(*40*) . The Mott density in single crystal CsPbBr$_3$ or MAPbBr$_3$ is $n_{Mott} \sim 8 \times 10^{17}$ cm$^{-3}$ (*40, 41*). Thus, $P_{th1}$ is far below $n_{Mott}$ while $P_{th2}$ is within a factor of 2-4 (depending on detuning) above $n_{Mott}$. The second lasing threshold may be related to the Bardeen-Cooper-Schrieffer (BCS) polariton lasing – a mechanism predicted theoretically (*42–44*) and more recently demonstrated experimentally in a GaAs quantum well microcavity(*39*). In the BCS mechanism, the e-h pair is no longer tightly bound as in the exciton, but Coulomb correlated to form a BCS-like pair (*39*) in the so-called non-degenerate electron-hole plasma (*40*). As the exciton density further increases and the e-h pair is no longer Coulomb correlated, the mechanism transitions to photonic lasing, i.e., stimulated emission from a degenerate electron-hole plasma (*40*). The gain mechanisms of both BCS polariton lasing and photonic lasing follow fermionic statistics, as opposed to bosonic statistics in the Bose-Einstein condensate (BEC) type polariton lasing(*39*). However, unambiguously distinguishing BCS polariton lasing from photonic lasing above the second threshold is challenging (*39*) and we leave this as a subject for future experimental and theoretical investigation.

In conventional semiconductor microcavities, stimulated emission from e-h plasmas usually comes from the same cavity mode that forms the polariton condensates. However, the multiple polariton modes of our samples give rise to distinct transition behavior, i.e., the stimulated emission occurs in the next lower energy cavity mode. This interpretation is supported by analysis of *P*-dependent lasing peak positions, FWHM, and dispersions. We focus on the cavity with $\delta = -15$ meV (see Fig S17 for $\delta = +1$ and $-40$ meV) and plot the *P*-dependent individual lasing peak intensities and FWHMs in Fig. 3D, and peak positions ($\lambda_P$) and total PL intensity in Fig. 3E. Above $P_{th1}$, the intensity of polariton lasing at ~533 nm rises rapidly by over two orders of magnitude in a small *P* window (1-4 ×$P_{th1}$), followed by a slow rise and a plateau. Above $P_{th2}$, the red-shifted stimulated emission lasing peaks at ~548 nm rise rapidly, by a similar rate as that of the polariton lasing peak. The nonlinear increases in lasing intensity above both thresholds are accompanied by blue shifts with increasing *P*, Fig. 3E. Above $P_{th1}$, repulsive polariton-polariton interaction induces the blue shift(*45*), while above $P_{th2}$, the blue shift can be attributed to cavity mode renormalization as a result of carrier-density-dependent reduction in the refractive index (*40*). The effect of a Mott transition is also evident in FWHM, open symbols in Fig. 3D. While the FWHM of the polariton lasing peak increases with *P*, the rate of increase clearly accelerates above $n_{Mott}$ and is similar to the rate of increase in FWHM for the stimulated emission lasing peak above $P_{th2}$. The appearance of the additional lasing peaks that are strongly red-shifted and mode hopped to the next lower lying cavity modes is consistent with the stimulated emission from e-h plasmas involving the emission of plasmons(*40*). An alternative interpretation of the two-threshold behavior is multimode polariton lasing, where polariton lasing may switch modes with increasing excitation density, giving rise to distinct thresholds due to the different relaxation dynamics(*46*). However, our observation of the second threshold being above $n_{Mott}$ and the likely transition into weak coupling disfavors this interpretation. For comparison, we also show in Fig. S18 results for a room temperature CsPbBr$_3$ microcavity, where only one lasing threshold near $n_{Mott}$ is observed, suggesting the transition to weak coupling across the threshold. Note that the FWHMs of the lower energy polariton lasing peaks also increases significantly when the exciton density exceeds $n_{Mott}$ (Fig. 3D, 3F) and are likely indicative of transition to stimulated emission.

Further evidence for polariton condensation comes from power-dependent dispersions, Fig. 4. At $P \sim 0.5 P_{th1}$, Fig. 4A$_1$, emissions from the two high energy spin-split polariton modes closest to $E_{ex}$ are the strongest at $k_\parallel = 0$ and decrease monotonically with increasing $k_\parallel$. By contrast, emissions from the lower energy polaritons with large detuning are confined to the bottleneck region at high $k_\parallel$. When $P$ is above $P_{th1}$ (Fig. 4A$_2$), emission occurs exclusively at $k_\parallel = 0$, accompanied by a stepwise rise in intensity, a dramatic peak narrowing, and a blue shift with increasing $P$ (see Fig. 3), as expected for exciton polariton condensation. When $P$ is increased to $\sim P_{th2}$, Fig. 4A$_3$, the lasing peak near the excitonic resonance is broadened and further blue-shifted. Additional red-shifted lasing peaks emerge and attributed to the photonic lasing mechanism accompanied by plasmon emission.

While the $\vec{B}$ fields do not determine the condensation dynamics, they are responsible for the pseudo-spins of the exciton-polaritons evolve as they undergo Bose scattering down the lower polariton branch and undergo phase transitions to the condensates. Below the condensation threshold, $\vec{B}_{XY}$ and $\vec{B}_T$ both contribute to give the spin-textures at finite $k_\parallel$, as exemplified near the diabolic point in Fig. 1H. As the system approaches condensation, population builds up around $k_\parallel = 0$ and the spin-texture evolves to one dominantly resulting from $\vec{B}_{XY}$, i.e., two orthogonally polarized polaritons illustrated in Fig. 1G. This is most obvious when the magnitude of detuning allows polariton condensation into a pair of spin-split modes with similar intensities, Fig. 4B.

Compared to conventional semiconductor microcavities, the orders of magnitude larger anisotropy in our samples can give rise to two competing condensates with different energies. The polarizations of the two condensates are determined by $\vec{B}_{XY}$, not the stochastic condensate polarizations in conventional microcavity exciton-polariton condensates(*47, 48*). The competing condensation dynamics can be understood from the balance between thermodynamics and kinetics. The lower energy anisotropic mode is thermodynamically favored, while the higher energy anisotropic mode is kinetically favored. The kinetic argument can be understood from the smaller curvature along the path towards minimum for the lower energy mode than that for the higher energy one, leading to a more efficient scattering pathway toward $k_{||}$ = 0 in the former(*49*). While the competition between the two spin polarized modes is always present due to the thermodynamic and kinetic factors, we find that modest detuning ($\delta \sim 20 - 40\ meV$) makes this competition most obvious in the simultaneous presence of both modes. Too small a detuning minimizes the effective field (discussion in Supplementary Text 1), and too large a detuning diminishes the difference in scattering pathways, leading to preferential condensation in the thermodynamically favored mode.

Our discovery of competing exciton-polariton condensates in anisotropic lead halide perovskite microcavities with artificial and tunable gauge fields offer exciting opportunities for the exploration and simulation of many-body SOC physics in the quantum fluid phase. While we demonstrate the gauge field and the interesting Rashba-Dresselhaus Hamiltonian for exciton-polaritons at finite $k_{||}$ near the diabolic point (Fig. 1), condensation (Figs. 2-4) occurs at $k_{||}$ = 0 where $\vec{B}_T$ = 0 and the effective magnetic field results exclusively from $\vec{B}_{XY}$. As an exciting future research direction, we may systematically tune the relative amplitude of $\vec{B}_T$ and, thus, the gauge field, and experimentally access momentum space away from $k_{||}$ = 0, particularly near the diabolic point. This is possible from off-normal and resonant optical excitation to launch propagating exciton-polaritons and condensates. Theory predicts that, for such propagating SOC condensates in a finite $k_{||}$ range above or below the diabolic point, the gauge field can result in dynamic instability, spin-textured phase separation, and strip formation in the quantum fluid(*10*). Our finding provides a starting point for the experimental exploration of a range of phenomena in SOC quantum fluids in a solid state and high-temperature model system, as done previously in cold atom based SOC-BECs(*19–21*).

# MATERIALS AND METHODS

**Microcavity fabrication**: All chemicals and regents were purchased from Sigma-Aldrich and used as received, unless noted otherwise. $CsPbBr_3$ precursor solution with a concentration of 0.4 M was prepared by dissolving stoichiometric 1:1 CsBr and $PbBr_2$ in dimethyl sulfoxide. $MAPbBr_3$ precursor solution with a concentration of 1.0 M was prepared by dissolving stoichiometric 1:1 MABr and $PbBr_2$ in dimethylformamide. Two types of distributed Bragg reflectors (DBRs) were used to fabricate the microcavities studied here. One type of DBRs consists of pairs of alternating silicon dioxide and silicon nitride, which were deposited on quartz or silicon wafers on an Oxford PlasmaPro PECVD NPG80 instrument. The center of the stop band was measured at 544 nm with a band width of 108 nm. The other type was custom produced by Thorlabs (subtractive magenta dichoric filter), and the center of the stop band was measured at 545 nm with a band width of 160 nm. Two opposing DBRs were bonded by epoxy, forming an empty cavity. Perovskite solutions were filled into the cavity through capillary action. The microcavity was then placed in a fused silica tube mounted in a single-zone Lindberg/Blue M tube furnace and connected to vacuum pump. The pressure in the tube was ~27 mTorr, the furnace temperature was ~80 °C. The solvent in the cavity slowly evaporated to form single-crystal perovskites within a few days. Representative optical images of $MAPbBr_3$ and $CsPbBr_3$ are shown in Fig. S1.

**Structural characterization**: The PXRD data were collected on as-grown samples using a Bruker D8 Advance powder X-ray diffractometer with Cu Kα radiation. The AFM measurements were performed using Bruker Dimension FastScan AFM.

**Spectroscopic characterization**: Angle-resolved PL measurements were carried out on a home-built confocal setup (see Fig. S5 for the details of the setup). The reflectance spectra of the crystals were measured with the same setup using a white lamp source. Low-temperature PL measurements were performed in a cryostat (Cryo Industries of America, RC102-CFM Microscopy Cryostat with LakeShore 325 Temperature Controller). The cryostat was operated at pressures <$10^{-7}$ mbar (pumped by a Varian turbo pump) and cooled with flow-through liquid nitrogen. The second harmonic of a Clark-MXR Impulse laser (repetition rate of 0.5 MHz, 250 fs pulses, 1040 nm), and a white light signal generated via $CaF_2$ and Impulse laser fundamental, was used to pump a home-built non-collinear optical parametric amplifier to generate 800 nm pulses which was used to generate 400 nm pulses via second harmonic generation. The beam size is expanded to ensure large-area illumination with a spot size ~25 μm and focused onto the sample with a far-field epifluorescence microscope (Olympus, IX73 inverted microscope) equipped with a × 40 objective with NA 0.6, with correction collar (Olympus LUCPLFLN40X) and a 490 nm long-pass dichroic mirror (Thorlabs, DMPL490R). The emission spectra were collected with a liquid nitrogen cooled CCD (Princeton Instruments, PyLoN 400B) coupled to a spectrograph (Princeton Instruments, Acton SP 2300i). We used the Lightfield software suite (Princeton Instruments) and LabVIEW (National Instruments) in data collection.

**Carrier density estimation.** We estimated the exciton density in photo-excited LHP single crystals using the following equation

$$n_{3D} = \frac{(1-R)(1-A_{DBR})A_{PVK}P}{f_{rep}\hbar\omega\pi r^2 d} \quad (6),$$

in which $n_{3D}$ is the exciton density, $R$ is the reflectivity of microcavity, $A_{DBR}$ is the absorbance of top DBR, $A_{PVK}$ is the absorbance of the perovskite crystal, $P$ is average power from the pulsed pump laser, $f_{rep}$ is the repetition rate of the pump laser, $\hbar\omega$ is the photon energy of the pump laser, $\pi r^2$ is the area of the pump spot, $d \sim 1$ μm is the thickness of the perovskite crystal (Fig. S3). Assuming negligible reflectivity and negligible absorbance of the top DBR, and unity absorbance of the perovskite crystal, the upper limit of the carrier density at the first threshold (~5.0 μJ cm$^{-2}$) is estimated to be ~$1.0 \times 10^{17}$ cm$^{-3}$, and the one at the second threshold (~53 μJ cm$^{-2}$) is ~$1.1 \times 10^{18}$ cm$^{-3}$.

We note that while the carrier density estimations in polariton systems may have large uncertainties, they can serve as an important check of the Mott transition. In our experiment, we observed unambiguous spectral signatures of the Mott transition above the second threshold: (1) a second threshold appears; (2) the photon lasing hops to the next, lower energy cavity modes due to plasmon emission; (3) the photon lasing transition is accompanied by significant linewidth broadening and apparent blueshift of the lasing peak, which are consistent with the disappearance of a bound excitonic state.

**Acknowledgements**

**Funding:** This work was solely supported by the Center of Programmable Quantum Materials (Pro-QM), an Energy Frontier Research Center of the US Department of Energy under grant number DE-SC0019443. The authors thank the generous help on sample preparation by Prof. Song Jin, Dr. Kwang Jae Lee, and Prof. Osman Bakr.

**Author contributions:**

X.-Y.Z., M.S.S., and Y.F. conceived this work. M.S.S. and Y.F. performed all optical experiments presented here. M.S.S. built the experimental setup, with assistance from A.P.S. and Y.D. Y.F. carried out sample fabrication, with inputs at the initial stage of this research project from D.H., M.D.S., and D.R.G.. Y.F., M.S.S, and X.-Y.Z. wrote the manuscript with inputs from all coauthors. X.-Y.Z. supervised the project. All authors participated in the discussion and interpretation of the results.

**Competing interests**: None.

**Data and materials availability**: All data needed to evaluate the conclusions in the paper are present in the paper or the Supplementary Materials.

**Supplementary Materials**

Supplementary text (1-4)
Supplementary Movie (S1 and S2)
Supplementary Figures (S1 to S20)

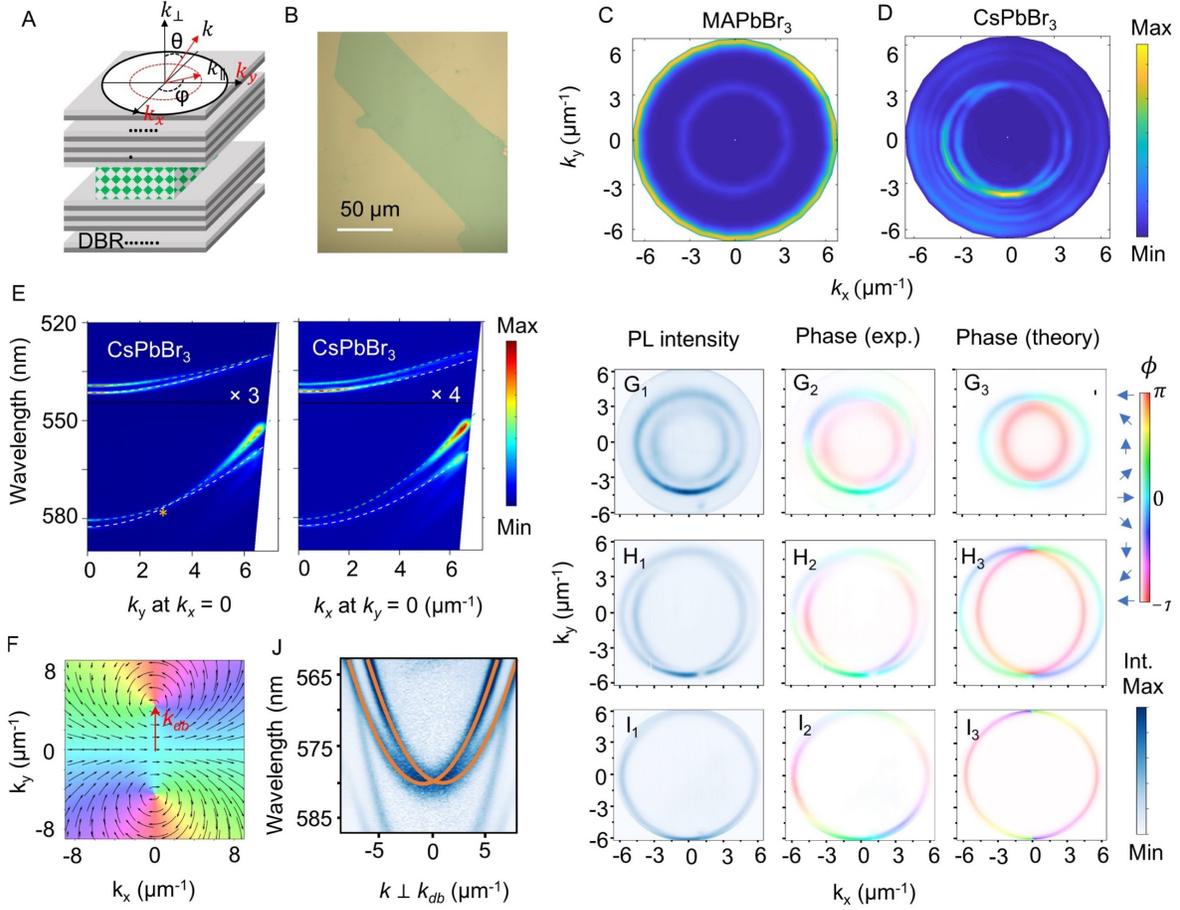

**Fig. 1. Spin textures in the anisotropic CsPbBr$_3$ perovskite microcavities.** (A) Schematic (left) and optical image of a CsPbBr$_3$ microcavity formed by two DBRs. The in-plane momentum is $k_\parallel = \frac{2\pi}{\lambda}\sin\theta$; $\theta$ is the polar angle of emission and $\lambda$ is the emission wavelength. $\varphi$ is azimuthal angle in the plane. (B) Optical image of a CsPbBr$_3$ single crystal grown in the microcavity channel. (C) Constant energy cross section at emission wavelength 565 nm from an isotropic MAPbBr$_3$ microcavity, showing a circular ring. (D) Constant energy cross section at 574.5 nm from an anisotropic CsPbBr$_3$ microcavity showing two offset circular modes. $k_x$ and $k_y$ are aligned with the <100> directions of the (pseudo-)cubic perovskite structures ($\phi_o = 0°$). (E) Dispersions along the $k_x$ and $k_y$ in the CsPbBr$_3$ microcavity, showing degeneracy at particular $k_y$ (marked by *) and anisotropic mode splitting at $k_\parallel = 0$. The dashed curves are theoretical SOC exciton polaritons from equation (4). (F) The momentum distribution of the effective magnetic field or gauge field and the Stokes polarization vectors parallel or anti-parallel to $\vec{B}_{eff}$. $k_{db}$ is the wavevector where the effective magnetic field vanishes. (G$_1$), (H$_1$), (I$_1$) show cuts of experimental dispersions at energies below (553.0 nm), at (548.9 nm), and above (544.7 nm) the diabolical points, respectively. (G$_2$), (H$_2$), (I$_2$) are the corresponding experimental spin-textures and (G$_3$), (H$_3$), (I$_3$) are the corresponding theoretical spin-textures from the Rashba-Dresselhaus Hamiltonian. (J) The energy dispersion intersecting the diabolical point in the direction perpendicular to $k_{db}$, showing two offset parabolas.

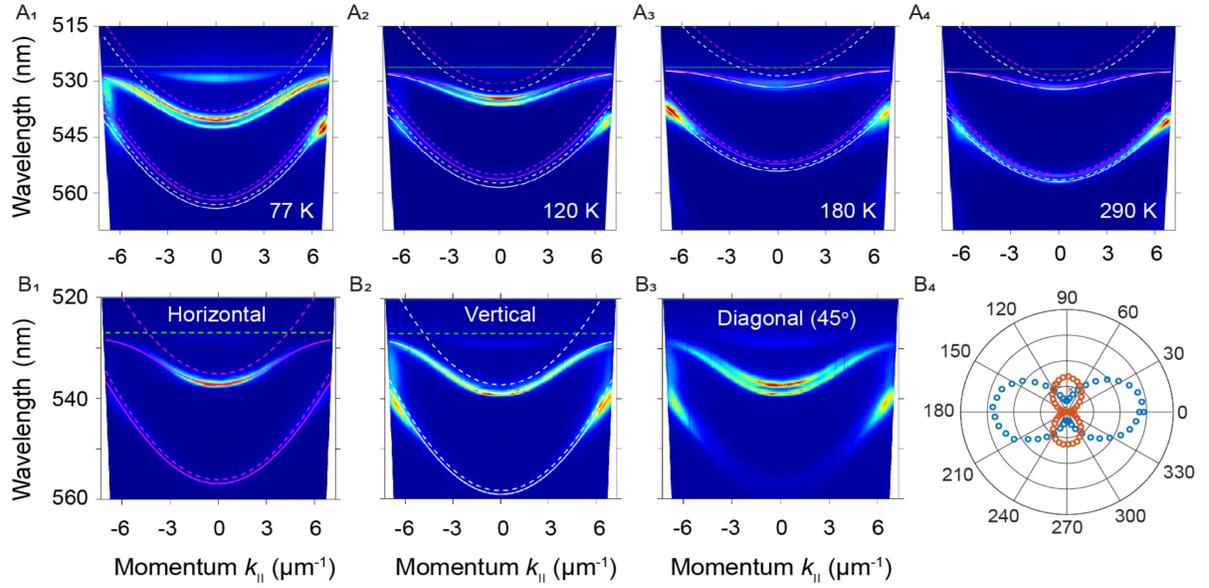

**Fig. 2. Temperature dependent and polarization resolved dispersions in CsPbBr$_3$ microcavities.** (A) Dispersion of a CsPbBr$_3$ microcavity measured at 77K (A$_1$), 120 K (A$_2$), 180 K (A$_3$), and 290 K (A$_4$). Dashed lines are the expected optical cavity dispersions; solid lines are the modeled polariton dispersions. The corresponding Rabi splittings are 25 ± 3, 25 ± 3, 18 ± 2, and 16 ± 2 meV, respectively. (B) Polarization-resolved dispersions of a CsPbBr$_3$ microcavity at 77 K. The polarization is (B$_1$) horizontal, (B$_2$) vertical, and (B$_3$) diagonal, with respect to the direction of the entrance slit in front of spectrometer. (B$_4$) Polarization-resolved PL emission of the two anisotropic modes ~538 nm at $k_\parallel = 0$, showing the two modes are mutually orthogonal, and linearly polarized.

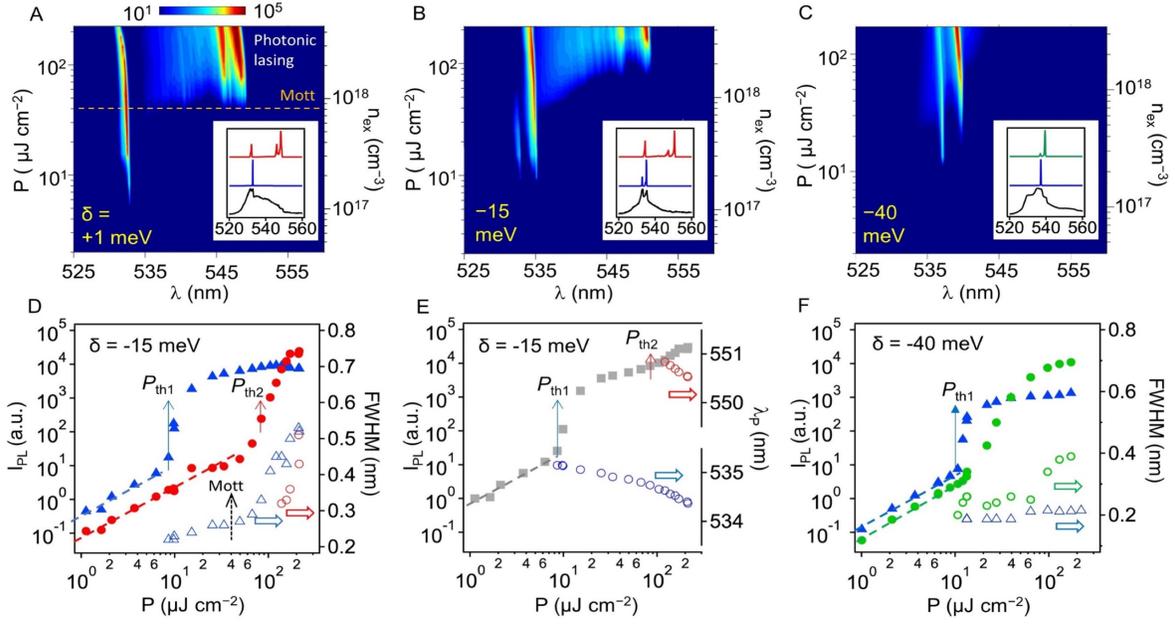

**Fig. 3. Condensation of SOC exciton-polaritons in CsPbBr$_3$ microcavities at 77 K.** (A) 2D pseudo-color plot of PL spectra as functions of pump fluence ($P$, left axis) or exciton density ($n_{ex}$, right axis) of three CsPbBr$_3$ microcavities with cavity detuning δ = (A) +1, (B) −15, and (C) −40 meV, respectively. The golden-dashed line in Fig. 3A marks the Mott density. Insets in A-C show normalized PL spectra at three pump fluences: (A) $P$ = 4.0 (black), 5.7 (blue), and 108 (red) µJ cm$^{-2}$; (B) $P$ = 1.6 (black), 9.7 (blue), and 170 (red) µJ cm$^{-2}$; (C) $P$ = 6.5 (black), 13 (blue), and 97 (green) µJ cm$^{-2}$. (D) Integrated PL intensity (left axis) as a function of $P$ in a log-log scale for the main lasing peaks (blue solid triangles, 531-536 nm; red solid circles, 545-552 nm), showing the two-threshold behavior for the δ = −15 meV cavity. Also shown are the FWHMs (right axis) of the lasing peaks at ~535 nm (blue open triangles) and ~550 nm (red open circles). (E) Total PL intensity (525-555 nm) (grey squares, left axis) and lasing peak positions (open circles, right axis) as a function of $P$ for the δ = −15 meV cavity. (F) $P$-dependences of integrated PL intensities (left axis) of the two lasing peaks from (blue solid triangles, 534-538 nm; green solid circles, 538-541 nm) and corresponding peak FWHMs (right axis; blue open triangles ~537 nm, green open circles, ~540 nm) for the δ = −40 meV cavity.

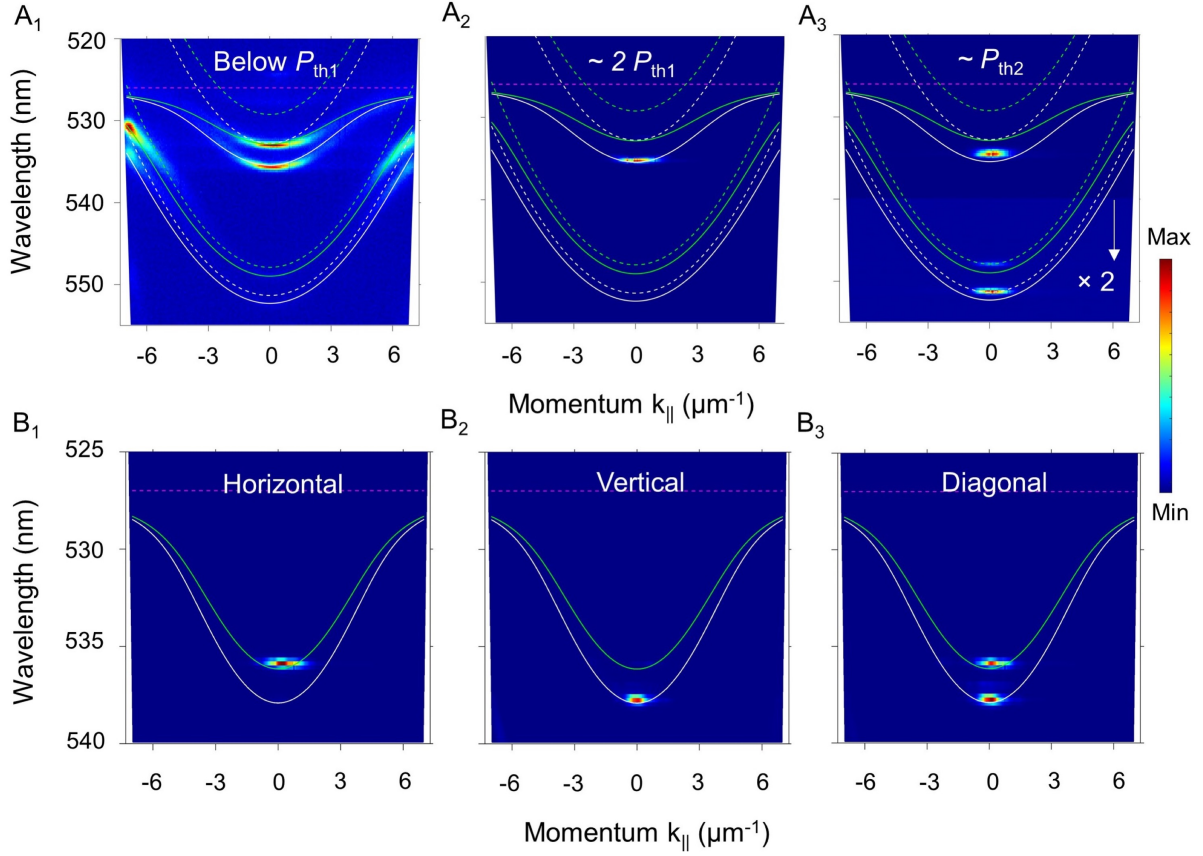

**Fig. 4. Competing polariton condensates with orthogonal polarizations at 77 K.** (A) Angle-resolved PL spectra of a CsPbBr$_3$ microcavity measured at (A$_1$) ~0.5$P_{th1}$, (A$_2$) ~2$P_{th1}$, and (A$_3$) ~$P_{th2}$. The cavity detunings are −18 and −98 meV for the higher energy (~534 nm) and lower energy (~550 nm) polariton modes, respectively. (B) Angle-resolved PL spectra of a CsPbBr$_3$ microcavity with the cavity detuning of −35 meV measured above $P_{th1}$ with the PL polarization along horizontal (B$_1$), vertical (B$_2$), and diagonal (B$_3$) direction with respect to the entrance slit of the spectrometer. The <100> axis of pseudo-cubic perovskite structure is aligned parallel to the entrance slit of the spectrometer. Dashed curves are the expected optical cavity dispersions; solid curves are the model polariton dispersions.

SUPPLEMENTARY INFORMATION

**Spin-Orbit Coupled Exciton-Polariton Condensates in Lead Halide Perovskites**

Michael S. Spencer[1] *, Yongping Fu[1] *, Andrew P. Schlaus[1], Doyk Hwang[1], Yanan Dai[1], Matthew D. Smith[2], Daniel R. Gamelin[2], and X.-Y. Zhu[1] †

[1] Department of Chemistry, Columbia University, New York, NY 10027, United States

[2] Department of Chemistry, University of Washington, Seattle, Washington 98195-1700, United States

*These authors contributed equally to this work.

†Author to whom correspondence should be addressed: xyzhu@columbia.edu

**Supplementary text 1:** *Justification of Hamiltonian*

The Hamiltonian is written using the formalism of a spin ½ particle in the presence of a magnetic field. This is possible due to the mathematical equivalence of the Poincarè Sphere and the Bloch Sphere, where the Poincarè Sphere represents the possible spin states of a free photon, whereas the Bloch Sphere represents possible spin states of an electron. To see why this is helpful, first consider relevant bases either in the presence of just the crystalline birefringence or just the usual TE-TM splitting that occurs in any DBR cavity

Let's consider first consider crystalline anisotropy. For the purposes of this discussion, we will consider only the in-plane components of the dielectric tensor, as this simplification is sufficient to capture the features observed in this report, although in principle at increasingly large angles of incidence, an incoming TM-polarized wave will be affected by this out of plane component of the dielectric function. Within this approximation, we can consider the in-plane birefringence to be that of a uniaxial crystal, where the orientation of the fast and slow axes with respect to the coordinate frame is determined by the angle the crystal is oriented. For example, if the a = (1,0,0) axis has the lowest refractive index, and is aligned along the x axis, then the y axis will correspond to b = (0,1,0), and would be the slow axis. In this case, an incident wave, oriented normal to the cavity, polarized along the y axis would experience an effective cavity length larger than one polarized along the x axis. Therefore, we will see two optical modes; the photon spin degeneracy is broken, in favor of modes polarized along either $\hat{x}$ or $\hat{y}$. The principle holds in general – any incident waves will be decomposed into the ordinary and extraordinary rays as a consequence of this in-plane birefringence, as the eigenmodes are polarized strictly along the crystallographic axes, which may be rotated as a consequence of crystal rotation.

TE-TM splitting, a more familiar consequence of DBR cavities, may also be incorporated. The physical origin of this effect is derived from the interfacial field matching conditions prescribed by Maxwell's Equations which distinguish between fields oscillating along or perpendicular to the plane of an interface. The TE wave will always be oscillating in the plane of the interface, whereas for increasingly large incidence angles the TM wave will oscillate perpendicular to the plane of the interface, thus giving it a different reflectivity. Near normal incidence, the polarization direction of a TM wave will also be in the plane, but will be polarized perpendicular to the TE polarization. It is also important to note that because the TE polarization direction is always perpendicular to the plane of incidence, the TE polarization direction will rotate $4\pi$ for a $2\pi$ rotation of polar angles. This effect is well documented (e.g. Panzarini et. al., 1999, Physics of the Solid State), and leads to a TE-TM splitting the grows like $k_\parallel^2$. Because the energy difference is with respect to the TE and TM polarizations, these polarizations remain the basis of the eigenmodes in a cavity where this is the dominant effect breaking spin symmetry.

Within the Poincarè Sphere formalism, the splittings discussed above may be easily incorporated as an effective field which leads to eigenfunctions that correspond to the proper basis. Importantly, it also allows for efficient calculation of a situation, such as in this report, where multiple effective fields are present, leading to optical modes which are not well described either in the basis of X,Y polarized modes or TE-TM polarized modes. In the same way that a magnetic field's direction can cause the electron spin to be aligned with that direction, we introduce an effective field which will cause the photon spin to be polarized along that given direction. For the fields considered here, that corresponds to all possible linearly polarizations, which lie on the equator of the sphere.

**Supplementary text 2:** *SOC polariton modeling*

When the SOC cavity photons are strongly coupled to an exciton resonance with Rabi splitting $\Omega$, we write the SOC polariton Hamiltonian, excluding the photonic and excitonic damping rates, as,

$$H_{Pol} = \begin{pmatrix} E_o + \dfrac{\hbar^2 k_\parallel^2}{2m} & \dfrac{\Omega}{2} & -\alpha e^{-i\phi_o} + \beta k_\parallel^2 e^{-2i\phi} & 0 \\ \dfrac{\Omega}{2} & E_{xc} & 0 & 0 \\ -\alpha e^{i\phi_o} + \beta k_\parallel^2 e^{2i\phi} & 0 & E_o + \dfrac{\hbar^2 k_\parallel^2}{2m} & \dfrac{\Omega}{2} \\ 0 & 0 & \dfrac{\Omega}{2} & E_{xc} \end{pmatrix}$$

This Hamiltonian is written with the assumption that the exciton resonance is itself isotropic, or at least negligible compared with the linewidth of the polariton modes. This Hamiltonian suggests four polariton bands: two lower polariton branches, and two upper polariton branches, where the two modes in a given branch correspond to one of two polarization states. To highlight the role of the effective magnetic field in the strong coupling regime, we use the eigenvectors of the traditional 2×2 coupled oscillator Hamiltonian to partially resolve the effects of strong coupling, and identify how strong coupling effectively modulates the strength of the effective magnetic field within the cavity. To achieve this, we perform a matrix transformation as demonstrated below:

$$H'_{Pol} = M^{-1} H_{Pol} M, \quad M = \mathbb{1} \otimes V$$

where $V$ is written by placing the eigenvectors of the coupled oscillator Hamiltonian as column vectors, i.e., we perform a unitary transformation into the eigenbasis of the isotropic polariton modes which have been introduced into the spin space via a tensor product. These eigenvectors are written below for completeness. The final matrix result (below) has also had the indices rearranged to highlight the re-appearance of the effective magnetic field, G, in this form of the matrix.

$$|v_1\rangle = X_{UP}|Exc\rangle + P_{UP}|Ph\rangle, \quad |v_2\rangle = X_{UP}|Exc\rangle + P_{UP}|Ph\rangle,$$

$$V = (v_1 \quad v_2) = \begin{pmatrix} X_{UP} & X_{LP} \\ P_{UP} & P_{LP} \end{pmatrix}$$

$$H'_{pol} = \begin{pmatrix} E_{UP} & P_{UP}^2 G & 0 & XPG \\ P_{UP}^2 G^* & E_{UP} & XPG & 0 \\ 0 & XPG^* & E_{LP} & P_{LP}^2 G \\ XPG^* & 0 & P_{LP}^2 G^* & E_{LP} \end{pmatrix}$$

Here we have written the effective magnetic field terms as $G = -\alpha e^{-i\phi_o} + \beta k_\parallel^2 e^{-2i\phi}$. The $X, P$ are the Hopfield Coefficients, which are given by: $X_{UP} = \frac{1}{\sqrt{2}}\left(1 + \frac{\Delta E}{\sqrt{\Delta E^2 + \Omega^2}}\right)^{\frac{1}{2}}$, $P_{UP} = \frac{1}{\sqrt{2}}\left(1 - \frac{\Delta E}{\sqrt{\Delta E^2 + \Omega^2}}\right)^{\frac{1}{2}}$, $\Delta E = E_{ex} - E_{ph} - \frac{\hbar^2 k_\parallel^2}{2m}$. The polariton branch energies are given by: $E_{UP,LP} = \frac{1}{2}\left(\Delta E \pm \sqrt{\Delta E^2 + \Omega^2}\right)$. Note that the Hopfield coefficients switch when considering a different polariton branch, i.e., $P_{UP} = -X_{LP}$, $X_{UP} = P_{LP}$. We maintain the distinction of the upper and lower branch terms to help clarify the role of the photonic component in the effective magnetic field. These terms quantify the projection of a given eigenvector onto the basis of the Hamiltonian – a pure photon or exciton state, and therefore their squares represent the photon or exciton fraction of a given eigenfunction. Their values can range anywhere from zero to one, and in the case of strong coupling at the anti-crossing point, they are both equal to $1/\sqrt{2}$.

The first important property to note about this transformed 4×4 Hamiltonian is that it can be well approximated as being two independent blocks, with a negligible coupling between the blocks. These coupling terms are given by $XPG$, and will couple a given polarization within the upper/lower branch to either polarization of the opposite polariton branch. The subscripts are dropped here because either subscript denotation would be equivalent. Because the coupling terms within a branch refer to a splitting amongst otherwise degenerate modes, they are much more consequential than coupling terms to another mode which is energetically well separated. The closest the upper and lower polariton branches can be is at the anti-crossing point, where they are separated by the Rabi splitting. In our system there is approximately 30 meV of separation. The effective magnetic field's energetic coupling in this same range is about an order of magnitude smaller, leading to a negligible coupling outside of a given polariton branch. As an example calculation, using typical values found within this report ($\Omega = 30\text{meV}, k_\parallel = 5\mu m^{-1}, \alpha = 5\ meV, \beta = 0.5\ meV\ (\mu m^{-1})^{-2}$), we identify that a typical effective magnetic field (restricted to the approximate form shown below) leads to approximately a 5 meV energy correction at the anti-crossing point, and that the addition of the full field correction, changes the energies less than 1 meV. For this reason, we further discuss the approximate form of the Hamiltonian;

$$H'_{pol} \approx \begin{pmatrix} E_{UP} & P_{up}^2 G & 0 & 0 \\ P_{up}^2 G^* & E_{UP} & 0 & 0 \\ 0 & 0 & E_{LP} & P_{LP}^2 G \\ 0 & 0 & P_{LP}^2 G^* & E_{LP} \end{pmatrix}$$

In this format, is now clear that the phase of the field field is approximately preserved, noting that there is some small polarization mixing we neglect in this form, although its total strength is strongly modulated by the appearance of strong coupling. To be precise, the strength of the effective magnetic field in a polariton branch is simply the photonic fraction at a given momentum times the effective field strength that corresponds to a purely photonic mode. Example simulations (e.g. Fig. S10) demonstrate that the anisotropy splitting is only important when the polariton mode exhibits an appreciable photonic component, which is why the lower polariton spin branches collapse onto one another as they approach the exciton resonance, whereas the upper polariton spin branches converge onto the purely photonic modes at high momenta. Fig. S10 further demonstrates the preservation of the effective magnetic field, where panels C,D show simulated spin texture at the energy corresponding to the diabolical point, showing that the major effect of strong coupling is to raise the effective mass of the mode.

For the modeling, we estimated the exciton resonance based on polariton dispersions with a series of cavity detunings and the reflectance spectra of the crystals. The wedged thickness of our sample allows us to continuously tune the cavity mode across the exciton resonance. By examining the polariton dispersions with various detunings, the exciton resonance can be estimated based on the flatten curvature at high $k$ momentum. For example, the exciton resonance at 77K is about 2.3525 eV, which is slightly larger than the PL emission peak. We have further confirmed that this value matched well to the absorption peak measured in ref. Nat. Communs. 2019, 10, 1175, i.e., 2.3525 eV at 80 K. Since our crystals have thickness about several micrometers, Fig. S3, the optical properties are expected to be same as bulk crystals measured in the reference. The reflectance spectra of our samples are also provided in Fig. S4. The excitonic feature at room temperature is much less noticeable than at low temperature. The various dispersions are provided in Fig. S13 for room temperature and Fig. S16 for 77 K.

**Supplementary text 3:** *Rashba-Dresselhaus Hamiltonian around the diabolical points*

In this section we provide a mathematical justification towards representing the Hamiltonian (Eq.1) as a modified Rashba-Dresselhaus Hamiltonian in the vicinity of the diabolical points. That is, we will expand this Hamiltonian for momentum vectors, $\mathbf{k}$, close to the of a diabolical point, $\mathbf{k}_{db}$, i.e., small $\mathbf{q} = \mathbf{k}_\parallel - \mathbf{k}_{db}$. We find it useful to decompose this small vector $\mathbf{q}$ into the components parallel and perpendicular to the diabolical point vectors – see figure S19 for visual aid. Further, for sufficiently small vectors $\mathbf{q}$ the projection onto $\mathbf{k}_\parallel$ is identical to the projection $\mathbf{k}_{db}$. By use of a small angle approximation and noting that $q_\perp$ is always perpendicular to the diabolical point vector $\mathbf{k}_{db} = \pm\sqrt{\frac{\alpha}{\beta}}\left(\cos\left(\frac{\phi_o}{2}\right), \sin\left(\frac{\phi_o}{2}\right)\right)^T$, we can write an equation that relates the change in the polar angle for small excursions about $\phi = \frac{\phi_o}{2}$: $\delta = \frac{\sin\left(\frac{\phi_o}{2}\right)q_x - \cos\left(\frac{\phi_o}{2}\right)q_y}{k_{db}} = \frac{q_\perp}{k_{db}}$.

We will prefer to keep things written in terms of $q_\perp, q_\parallel$ for convenience, although to appreciate the connection to the Rashba-Dresselhaus Hamiltonian, they will be decomposed into $q_x, q_y$ later. We also note that the quantity, $k_\parallel^2$, which appears repeatedly in Eq.1, will now be represented in terms of the diabolical point and the small vector q: $k_\parallel^2 = k_{db}^2 + q^2 + 2\mathbf{k}_{db} \cdot \mathbf{q} = k_{db}^2 + q^2 + 2k_{db}q_\parallel$.

Using small angle approximation, we write the exponential (Eq.1) as follows:

$$e^{-2i\phi} = e^{-2i\left(\frac{\phi_o}{2}+\delta\right)} \approx \cos(\phi_o) + \frac{2q_\perp}{k_{db}}\sin(\phi_o) - i\left(\sin(\phi_o) - \frac{2q_\perp}{k_{db}}\cos(\phi_o)\right)$$

Using this expression, we can express the upper right hand side of the Hamiltonian as:

$$H_{0,1} = -\alpha\big(\cos(\phi_o) - i\sin(\phi_o)\big) + \big(\alpha + \beta q^2 + 2\sqrt{\alpha\beta}q_\parallel\big)\left(\cos(\phi_o) + \frac{2q_\perp}{k_{db}}\sin(\phi_o) - i\left(\sin(\phi_o) - \frac{2q_\perp}{k_{db}}\cos(\phi_o)\right)\right)$$

After cancelling certain terms, we are left with the following expression:

$$H_{0,1} = \beta q^2\big(\cos(\phi_o) - i\sin(\phi_o)\big) + \alpha\frac{2q_\perp}{k_{db}}\big(\sin(\phi_o) + i\cos(\phi_o)\big) + 2\sqrt{\alpha\beta}q_\parallel\big(\cos(\phi_o) - i\sin(\phi_o)\big)$$

$$H_{0,1} = \left(\beta q^2 + i\alpha\frac{2q_\perp}{k_{db}} + 2\sqrt{\alpha\beta}q_\parallel\right)e^{-i\phi_o} = \left(\beta q^2 + 2\sqrt{\alpha\beta}(q_\parallel + iq_\perp)\right)e^{-i\phi_o}$$

By similar calculation, the Hamiltonian can be written in the following form:

$$H = \begin{pmatrix} E_o + \frac{\hbar^2}{2m}(k_{db}^2 + q^2 + 2k_{db}q_\parallel) & (\beta q^2 + 2\sqrt{\alpha\beta}(q_\parallel + iq_\perp))e^{-i\phi_o} \\ (\beta q^2 + 2\sqrt{\alpha\beta}(q_\parallel - iq_\perp))e^{i\phi_o} & E_o + \frac{\hbar^2}{2m}(k_{db}^2 + q^2 + 2k_{db}q_\parallel) \end{pmatrix}$$

$$H = \left(E_o + \frac{\hbar^2 k_\parallel}{2m}\right)\mathbb{1} + \begin{pmatrix} 0 & \beta q^2 + 2\sqrt{\alpha\beta}(q_\parallel + iq_\perp) \\ \beta q^2 + 2\sqrt{\alpha\beta}(q_\parallel - iq_\perp) & 0 \end{pmatrix} = H_o + H_I$$

In the second line, we have identified a distinction between the diagonal terms and the off-diagonal terms, which correspond to the field, that dictate the phase of the polariton branches at a given wavevector. We have neglected at this point the exponential phase terms, which correspond to a rotation about the z-axis on the Poincare sphere, i.e., $H'_I = R_z(\phi_o) H_I R_z(-\phi_o)$. This corresponds to a universal shift of the emission angle. It is not physically observable, but in principle is derived from the rotation of the crystal by an angle $\phi_o$.

At this point we switch to representing the small wavevector in terms of its x and y components.

$$q_\parallel = q \cdot \frac{k_\parallel}{|k_\parallel|} \approx \cos\left(\frac{\phi_o}{2}\right)q_x + \sin\left(\frac{\phi_o}{2}\right)q_y; \quad q_\perp = q \cdot \frac{-k_\parallel \times z}{|k_\parallel|} \approx \sin\left(\frac{\phi_o}{2}\right)q_x - \cos\left(\frac{\phi_o}{2}\right)q_y$$

$$H_I = \beta q^2 \sigma_x + \begin{pmatrix} 0 & \kappa_x q_x + \kappa_y q_y + i\kappa_y q_x - i\kappa_x q_y \\ \kappa_x q_x + \kappa_y q_y - i\kappa_y q_x + i\kappa_x q_y & 0 \end{pmatrix}$$

Here we have used $\kappa_x = 2\sqrt{\alpha\beta}\cos\left(\frac{\phi_o}{2}\right), \kappa_y = 2\sqrt{\alpha\beta}\cos\left(\frac{\phi_o}{2}\right)$ for notational simplicity. In this way we finally write

$$H = \left(E_o + \frac{\hbar^2}{2m}(k_{db}^2 + q^2 + k_{db}q_\parallel)\right)\mathbb{1} + \beta q^2 \sigma_x + \kappa_x \boldsymbol{\sigma} \cdot \boldsymbol{q} + \kappa_y \boldsymbol{\sigma} \times \boldsymbol{q}$$

This is the form of the Hamiltonian about the diabolical points, in the limit of very small $\boldsymbol{q}$. Here when $\boldsymbol{\sigma}$ is the vector of Pauli matrices. Because of the cavity the momentum in the z-direction is quantized, we consider only the x,y directions for the momenta and Pauli matrices. Similarly, we consider only the z-component of $\boldsymbol{\sigma} \times \boldsymbol{q}$. We can observe now that the Hamiltonian takes on the form of a modified Rashba-Dresselhaus Hamiltonian, with the addition of a term that goes like $\beta q^2$. Importantly, note that the weighting of the Rashba and Dresselhaus terms correspond to the crystal rotational angle, i.e., that for certain crystal orientations a purely Rashba- or Dresselhaus-like Hamiltonian can be achieved.

We now will highlight the connection of this interaction Hamiltonian with the graphene tight-binding Hamiltonian similarly expanded about the Dirac cone region. We can write the Hamiltonian in the following way, representing the phase terms as a complex exponential, rather than the $\kappa_x, \kappa_y$ terms:

$$H_I \approx \beta q^2 \boldsymbol{\sigma}_x + 2\sqrt{\alpha\beta}\begin{pmatrix} 0 & (q_x - iq_y)e^{\frac{i\phi_o}{2}} \\ (q_x + iq_y)e^{-\frac{i\phi_o}{2}} & 0 \end{pmatrix}$$

This is the same Hamiltonian, but now written in a way that more intuitively relates to the crystal rotation angle. We can now clearly see how the rotation of the crystal rotates the momentum vector providing the coupling between the previously spin-degenerate branches. In addition, we note that this is very similar to the Hamiltonian in graphene (to first order in $q$, see Neto, A.C., Guinea, F., Peres, N.M., Novoselov, K.S. and Geim, A.K., 2009. *Reviews of Modern Physics*, *81*(1), p.109), with the addition of a 'twist' which can turn the spin texture from a divergence to a curl structure. These properties are displayed in figure S.20, which shows the evolution of the spin texture upon crystal rotation.

We will now turn our attention to the eigenvalue within this same region. Based on the form of our Hamiltonian, we should expect to observe the traditional spin-split parabolas characteristic of a Rashba-Dresselhaus type Hamiltonian. We employ the same type of expansion of the eigenvalue equation to arrive at the following expression:

$$E_\pm(q \approx 0) - \left(E_o + \frac{\hbar^2 k_{db}^2}{2m}\right) = \frac{\hbar^2}{2m}(q^2 + 2\boldsymbol{k}_{db} \cdot \boldsymbol{q}) \pm 2\sqrt{\alpha\beta}q$$

If we write the eigenvalue strictly along the direction $q_\perp$, we then have the familiar Rashba-Dresselhaus eigenvalue equation:

$$E_\pm(q \approx 0) - \left(E_o + \frac{\hbar^2 k_{db}^2}{2m}\right) = \frac{\hbar^2}{2m}q_\perp^2 \pm 2\sqrt{\alpha\beta}q_\perp$$

**Supplementary text 4:** *Stokes Vectors and Connecting Hamiltonian To Polarization*

*Stokes Vector Experimental Characterization:* Stokes Vector is a 4-dimensional vector, which characterizes the full polarization state of an optical state, projecting the polarization onto 3 different bases, with a fourth component describing total intensity/ degree of polarization.

$$S = \begin{pmatrix} S_o \\ S_1 \\ S_2 \\ S_3 \end{pmatrix}$$

Here, $S_o$ is the total intensity, and the degree of polarization is given by:

$$p = \frac{\sqrt{S_1^2 + S_2^2 + S_3^2}}{S_o}$$

$S_1, S_2, S_3$ measure the projection onto the basis of horizontal/vertical polarizations, diagonal/anti-diagonal polarizations, and circular polarizations, respectively. So-called Mueller Matrices, $M$ can be constructed which allow computation of an output stokes vector, given an input stokes vector and an optical element;

$$\mathbf{S'} = \mathbf{MS}$$

For the elements of interest here (quarter wave plate at some angle $\theta$, and linear polarizer fixed to horizontal polarization transmission) we have the Mueller Matrices:

$$M_{hlp} = \frac{1}{2}\begin{pmatrix} 1 & -1 & 0 & 0 \\ -1 & 1 & 0 & 0 \\ 0 & 0 & 0 & 0 \\ 0 & 0 & 0 & 0 \end{pmatrix}$$

$$M_{qwp} = \begin{pmatrix} 1 & 0 & 0 & 0 \\ 0 & \cos^2(2\theta) & \cos(2\theta)\sin(2\theta) & \sin(2\theta) \\ 0 & \cos(2\theta)\sin(2\theta) & \sin^2(2\theta) & -\cos(2\theta) \\ 0 & -\sin(\theta) & \cos(2\theta) & 0 \end{pmatrix} \quad ($$

From these we can compute the $S'_o$ measured by the camera, after passing through both optical elements, in terms of the original stokes vector emitted from the microcavity:

$$S'_o = \frac{1}{2}(S_o + S_1\cos^2(2\theta) + S_2\cos(2\theta)\sin(2\theta) + S_3\sin(2\theta))$$

To determine the Stokes vector, we measure the energy-, momentum-dependent dispersion for many choices of $\theta$, and then perform a discrete Fourier transform on this data to extract the Stokes vector parameters. Often the data is presented showing the angle of the linear polarization, which corresponds to the angle along the equator of the Poincaré Sphere, which may be computed as $\phi_{stokes} = \tan^{-1}(\frac{S_2}{S_1})$. Note that this angle is not the same as the geometric angle expressed in the Hamiltonian. Rather, it is the same angle as those labelled in the scale bar of the phase plots, either experimental (Figs. 1G$_2$, 1H$_2$ & 1I$_2$) or theoretical (Figs. 1G$_3$, 1H$_3$ & 1I$_3$).

*Connecting Hamiltonian To Polarization:* The Hamiltonian under consideration here is written within the basis of circular polarizations, and its eigenvectors are complex-valued superpositions of left- and right-handed circular polarization states:

$$|\psi\rangle = c_1|+\rangle + c_2|-\rangle$$

From this (normalized) state, one can calculate the Stokes vector parameters, based on the transformation between the basis states;

$$\mathbf{S} = \begin{pmatrix} |c_1|^2 + |c_2|^2 \\ 2 * \Re(c_2^* c_1) \\ -2 * \Im(c_2^* c_1) \\ |c_1|^2 - |c_2|^2 \end{pmatrix}$$

where $\Re$, $\Im$ denote taking only the real or imaginary part of an expression. As noted before, this allows for the computation of the in-plane angle of the Stokes vector, i.e., the orientation of the polarization: $\phi_{stokes} = \tan^{-1}\left(\frac{S_2}{S_1}\right) = \tan^{-1}\left(-\frac{\Im(c_2^* c_1)}{\Re(c_2^* c_1)}\right)$. Results of these calculations are shown in Fig.S10, showing the spin texture of an entire mode, and are used to generate the theoretical comparison in Figs. 1G$_3$, 1H$_3$ & 1I$_3$ and Figs S10, S20 by displaying the hue from the phase calculation and in the case of figures S10, the saturation is derived from using a Lorentzian-convolution of eigenvalue dispersion and then taking a cut of that convolution at a given energy.

**Supplementary movies**

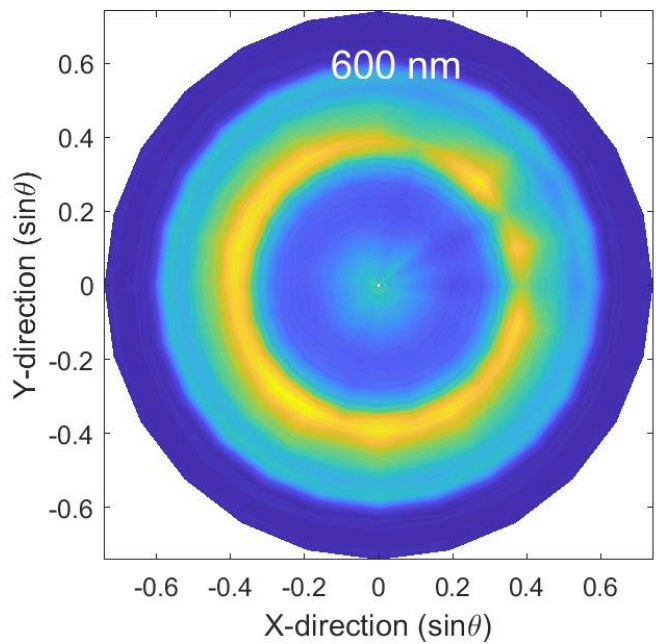

**Movie S1.** Wavelength resolved Fourier space imaging of PL from an isotropic MAPbBr3 microcavity at room temperature.

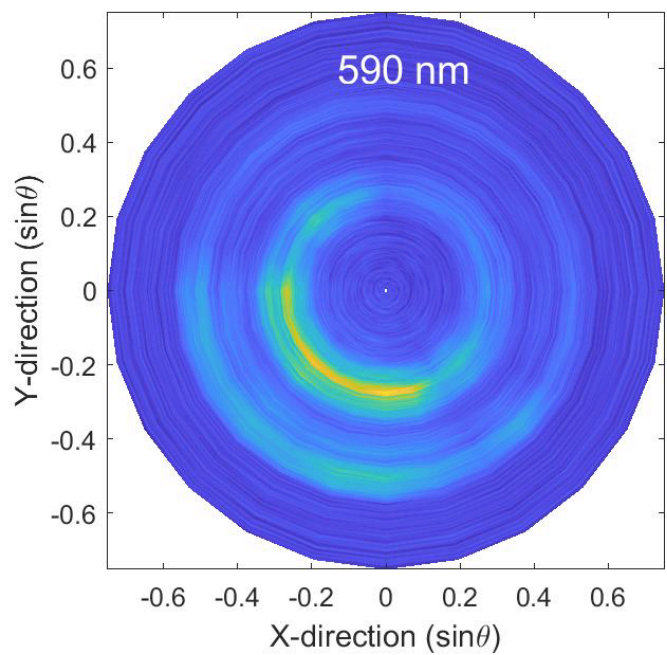

**Movie S2.** Wavelength resolved Fourier space imaging of PL from an anisotropic CsPbBr3 microcavity at room temperature.

**Supplementary figures**

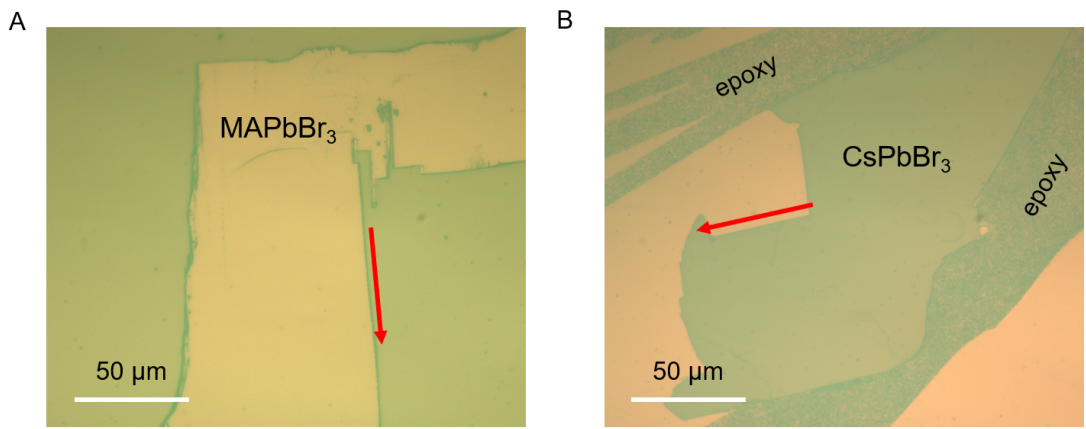

**Fig. S1. Optical image of representative MAPbBr$_3$ (A) and CsPbBr$_3$ (B) single crystals in microcavity.** Based on the crystal shape and the crystal growth behavior, we identify the <100> axis of the (pseudo)cubic perovskite structure (indicated by the red arrows). At room temperature, the MAPbBr$_3$ is cubic phase (space group $Pm\bar{3}m$, $a$ = 5.948 Å), whereas the CsPbBr$_3$ is orthorhombic phase (space group $Pbnm$, $a$ = 8.202 Å, $b$ = 8.244 Å, $c$ = 11.748 Å). The orthorhombic phase can be considered as a pseudo-cubic phase, in which the unit cell parameter ac can be related to that of the orthorhombic cell, by $a_c = a/\sqrt{2} \approx b/\sqrt{2} \approx b/2$. Note that the spot-like area in B is the epoxy used to bond the DBRs.

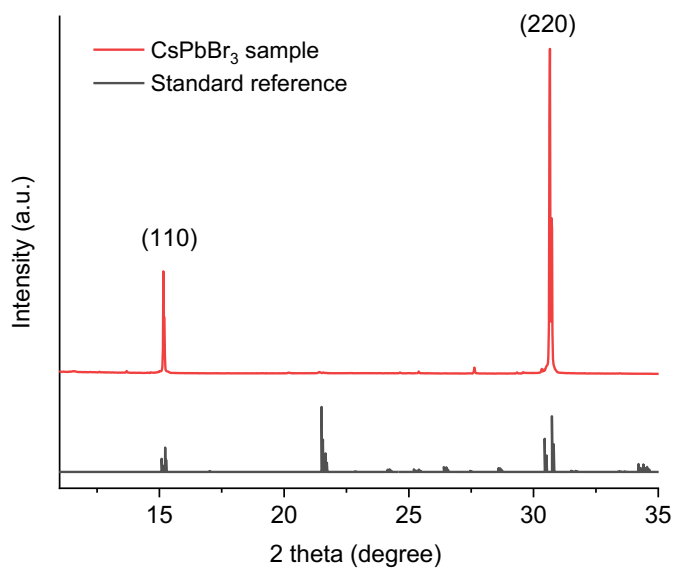

**Fig. S2. PXRD of the as-grown CsPbBr$_3$, in comparison with the standard PXRD pattern for CsPbBr$_3$ perovskite phase at room temperature**. The perovskite crystals are highly oriented along <110> direction in the z direction.

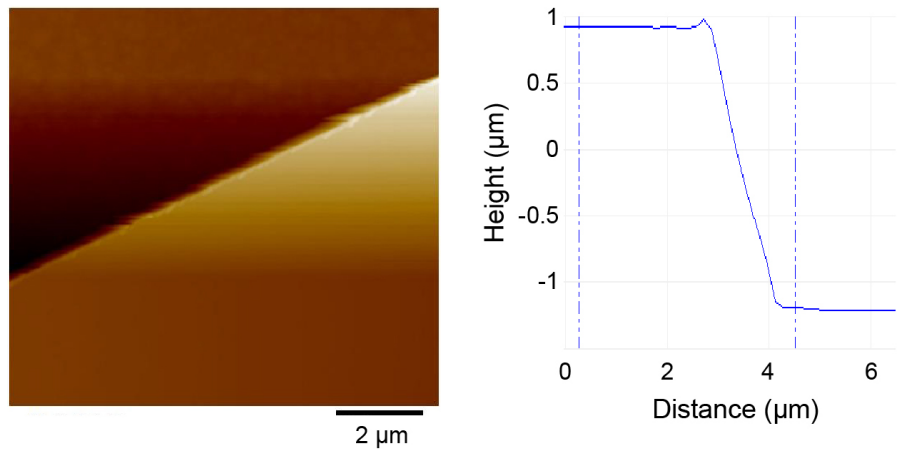

**Fig. S3. Atomic force microscopy of a representative CsPbBr$_3$ crystal, showing a thickness of ~2 μm.**

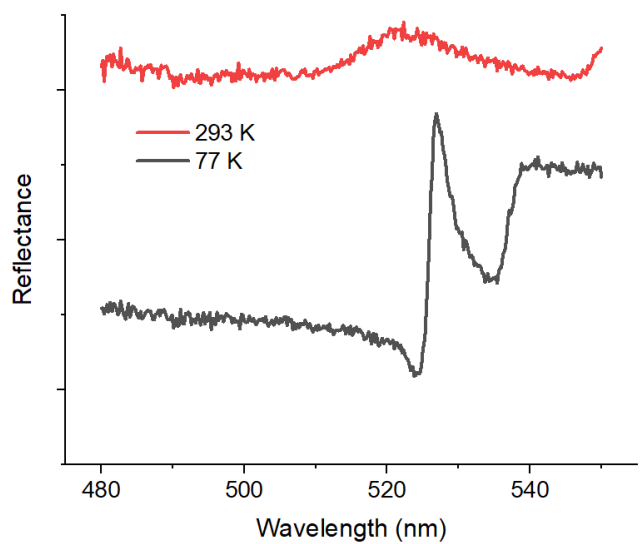

**Fig. S4. Reflectance spectra of the CsPbBr$_3$ crystal measured at room temperature and 77 K.**

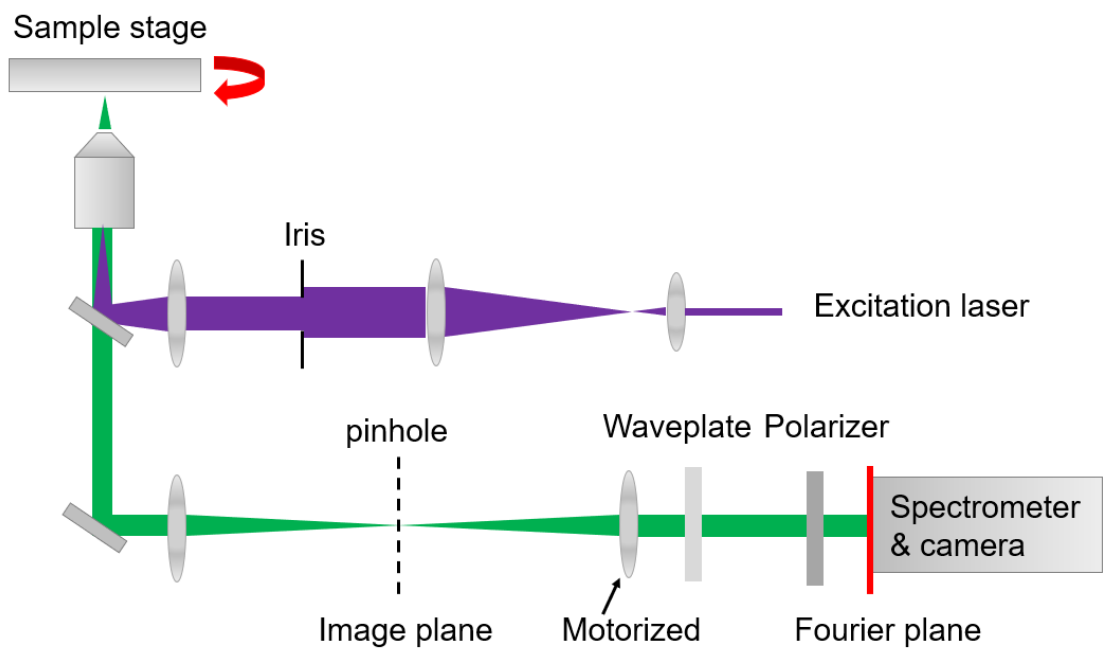

**Fig. S5**. Optical setup of Fourier space imaging with polarization-resolved capability. The pump laser is focused near the back focal plane of the objective lens to obtain a large uniform excitation spot on the sample. The Fourier plane is imaged through the entrance slit of the spectrometer, where the grating disperses the wavelengths perpendicular to the orientation of the entrance slit, onto a 2D camera array. The fully energy-momentum resolved Fourier space imaging can be constructed by (i) rotating the perovskite microcavity, i.e., equivalent to rotating the Fourier plane; or (ii) scanning the lens, i.e., equivalent to moving the Fourier plane across the entrance slit of the spectrometer. The full Stokes vectors can be measured through the use of a rotating quarter waveplate and a linear polarizer positioned before the entrance slit. The emission polarization is measured by positioning a half waveplate and a linear polarizer in front of the entrance slit.

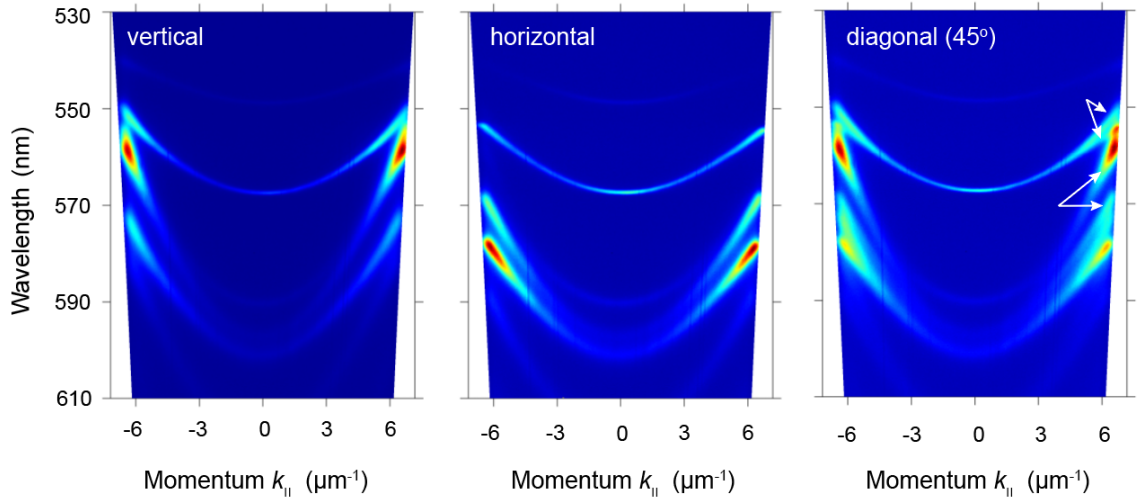

**Fig. S6. Polarization-dependent energy dispersions of an isotropic MAPbBr$_3$ microcavity, with the <100> axis of the perovskite crystals set to be parallel to the entrance slit of the spectrometer**. The sample is at room temperature. The linear polarizer is oriented (i) vertical, i.e., parallel to the direction of the entrance slit; (ii) horizontal, i.e., perpendicular to the entrance slit; (iii) diagonal, i.e., oriented 45º with respect to the entrance slit. One can see that the TE and TM modes are degenerate at $k_\parallel = 0$. However, energy splitting between the two modes becomes evident at high $k_\parallel$, as shown by the arrows in c pointing two pairs of TE-TM modes. In general, the TE-TM splitting occurs when the frequency of the cavity is not equal to the center frequency of the stop bands in the DBRs, and the magnitude of the splitting scales with $k_\parallel^2$.

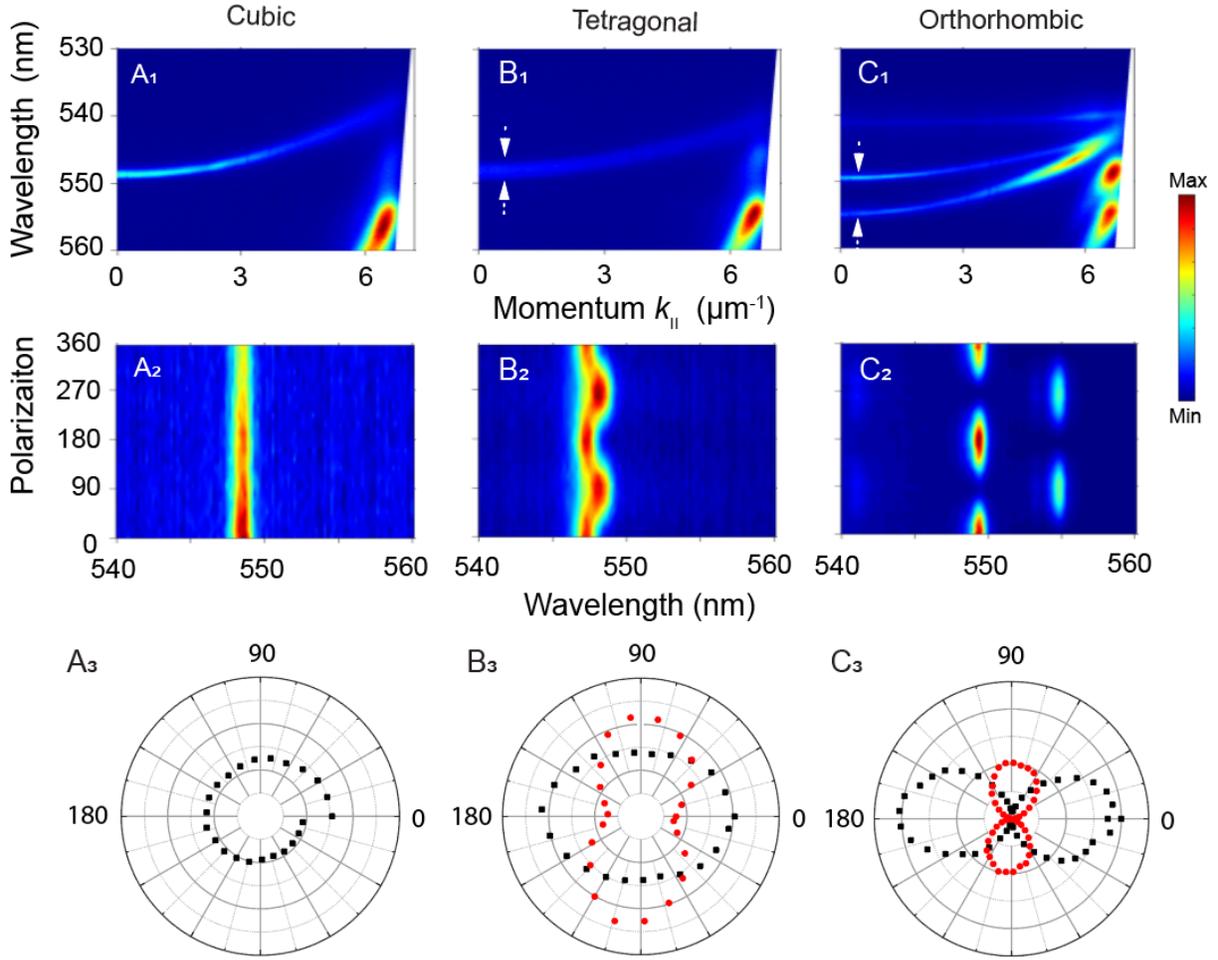

**Fig. S7. Tuning the synthetic SOC Hamiltonian by way of structural phase transition in a MAPbBr$_3$ microcavity:** (A) cubic phase; (B) tetragonal phase; (C) orthorhombic phase. Top panels: energy dispersions of the three phases. The cubic phase exhibits single mode dispersions at $k_\parallel = 0$. By comparison, the tetragonal phase exhibits a small mode splitting at $k_\parallel = 0$, due to small in-plane anisotropy. As the structure enters the orthorhombic phase, the mode splitting at $k_\parallel = 0$ becomes obvious, due to the increased in-plane anisotropy. Middle panels: polarization-resolved PL spectra at $k_\parallel = 0$ of the three phases. The cubic phase exhibits no polarization dependence at $k_\parallel = 0$, as expected in the absence of crystalline anisotropy. Note that the intensity gradually decreases due to sample degradation. By comparison, the tetragonal and orthorhombic phases show the two split modes are orthogonal to each other. Bottom panels: polarization-resolved PL emissions of the peaks at $k_\parallel = 0$. One can observe the polarization pattern changes as the structure undergoes phase transition. Note that the discontinuity at 0° is due to slight photobleaching of the sample during the time course of the measurement.

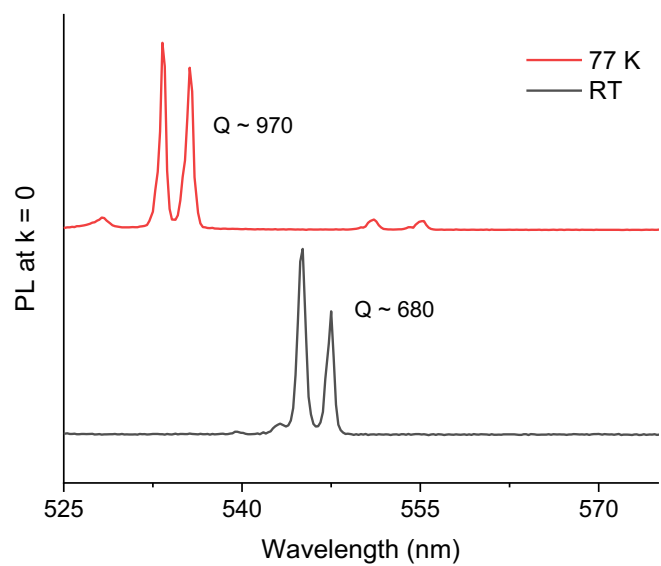

**Fig. S8**. **PL emission of the polariton at k = 0 from representative microcavities measured at room temperature and 77 K.**

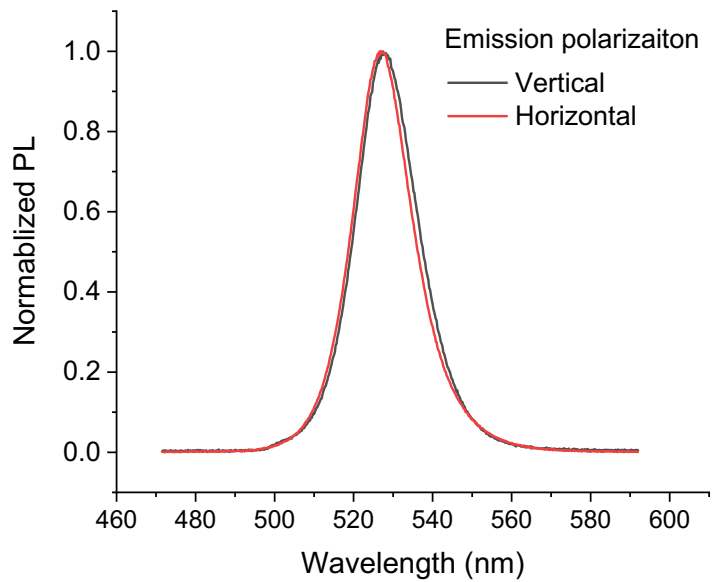

**Fig. S9**. **Polarization-resolve photoluminescence spectra of a CsPbBr$_3$ single crystal at room temperature.** One of the <100> axis of the (pseudo)cubic crystal is aligned with the entrance slit in front of the spectrometer. The vertical and horizontal polarizations are relative to the entrance slit.

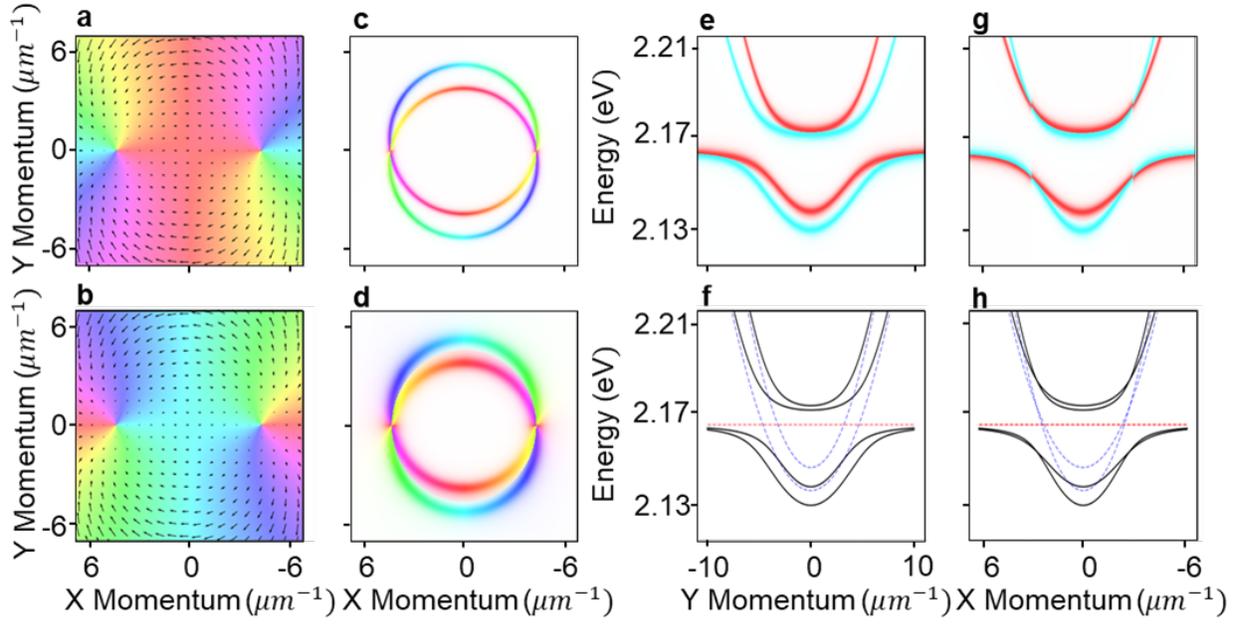

**Fig. S10. Phase-Resolved Properties of Anisotropic Polariton Hamiltonian (a,b).** The hue represents the stokes vector orientation along the equator of the Poincarè Sphere, and the vector represents the effective field orientation and magnitude at a given momentum vector. Note that for eigenvector (**a**) the phase is aligned with the field, whereas for the other eigenvector (**b**) the phase is opposite to the effective field. (**c,d**) Figure **c** represents the phase-resolved simulated emission intensity about the diabolical point energy, in the absence of strong coupling. Figure **d** represents the phase-resolved simulated emission intensity about the diabolical point energy, including the strong coupling, showing the effective increase in the emission linewidth as a function of the decreased slope. (**e,f**) Black lines show the result of the full 4x4 Hamiltonian. The dotted red line shows the exciton energy, and the dotted blue lines show the results of the anisotropy Hamiltonian, before accounting for strong coupling. (**g,h**) Simulated phase-resolved momentum cuts depict a switch in the emission polarization of the polariton mode along the x direction (**h**), and showing the convergence of the two different polarized polariton modes as the exciton resonance is approached, along the y direction (**g**). All calculations presented here use the band parameters: $E_o = 2.1415\,eV$, $m = 2.4 * 10^{-4} m_e$, $\alpha = 5 * 10^{-3}\,eV$, $\beta = 2.5 * 10^{-4}\,eV \cdot \mu m^{-2}$, $E_{xc} = 2.165\,eV$

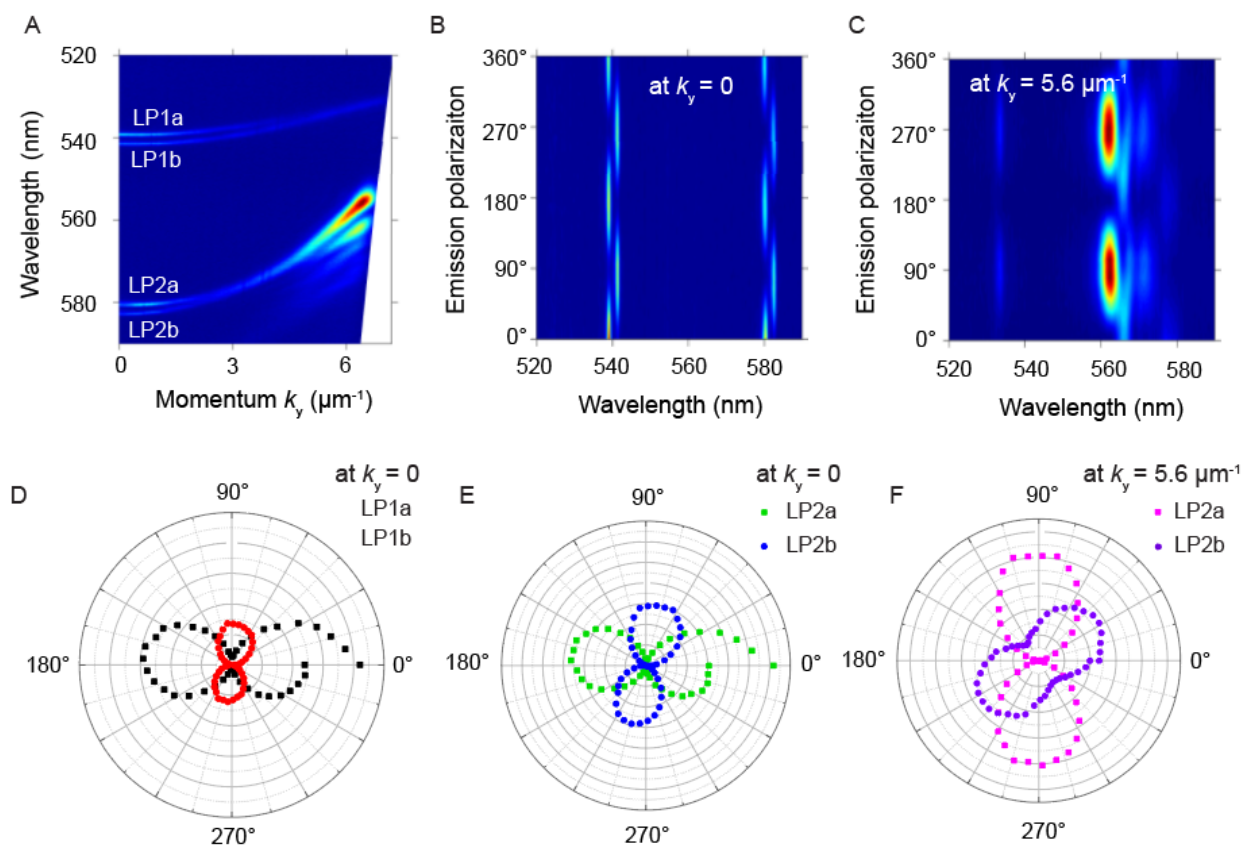

**Fig. S11. Polarization-resolved energy dispersion along $k_y$ in CsPbBr$_3$ microcavity measured at room temperature.** (A) The energy dispersion along $k_y$. The two pairs of anisotropic modes of interest here are labeled as LP1a, LP1b, LP2a, and LP2b. (B, D, E) Polarization-resolved PL emission at $k_y = 0$, showing the two modes in a given pair are mutually orthogonal, and linearly polarized, in agreement with the theoretical predictions shown in Fig. S5. (C, F) Polarization-resolved PL emission of the LP2a at $k_y \sim 5.6$ µm$^{-1}$, above the diabolical point, showing linear polarization orthogonal to that of at $k_y = 0$. This is consistent with the pseudospin singularity at the diabolical point shown in Fig. S5. At high momentum, the emission of LP2b mode severely mix with the emission from the next, energetically lower mode or the nearby Bragg mode of the cavity, causing deviations from the theoretical predictions. Note that the discontinuity at 0º is due to slight photobleaching of the sample during the time course of the measurement.

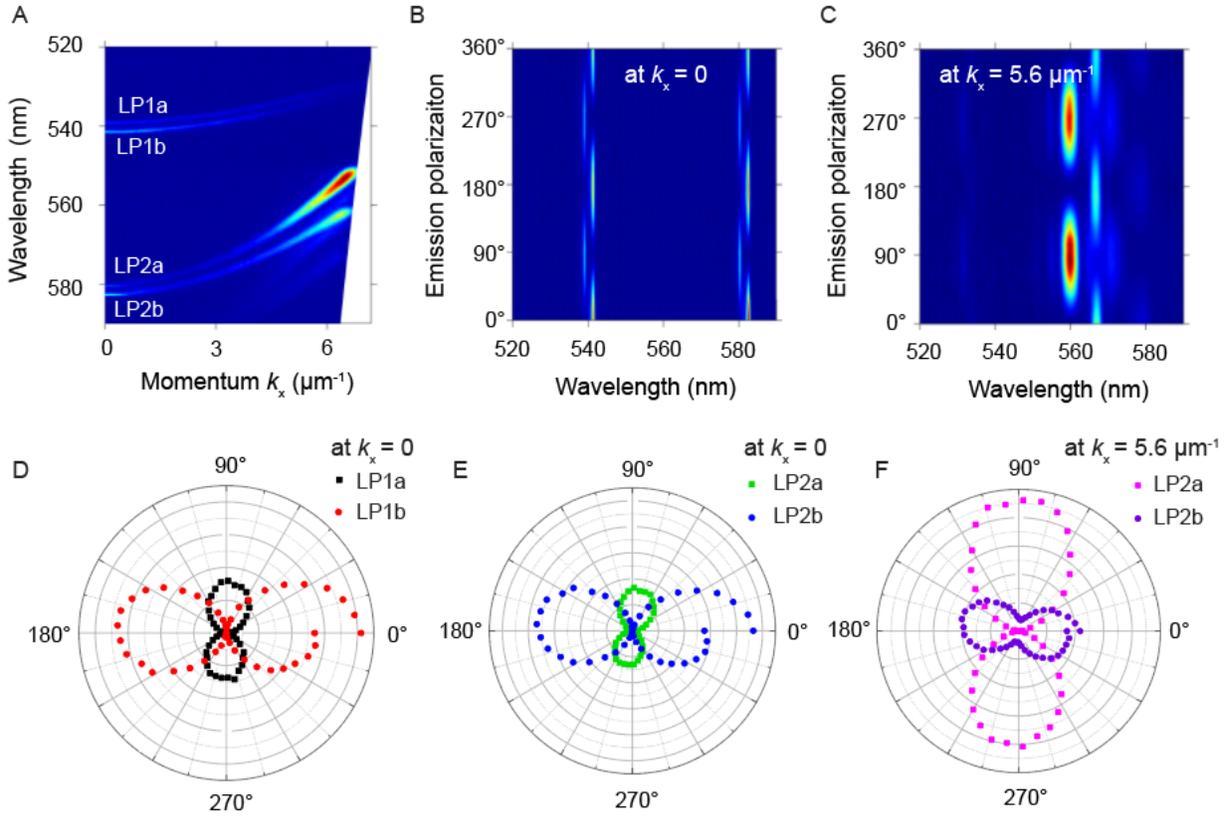

**Fig. S12. Polarization-resolved energy dispersion along $k_x$ in an anisotropic CsPbBr$_3$ microcavity, measured at room temperature.** (A) The energy dispersion along $k_x$. (B, D, E) Polarization-resolved PL emission at $k_x = 0$, showing the two modes in a given pair are mutually orthogonal linear polarized, in agreement with the theoretical predictions shown in Fig. S5. (C, F) Polarization-resolved PL emission at $k_x \sim 5.6$ μm$^{-1}$, showing roughly orthogonal linear polarizations for the two modes. The polarizations of the two modes maintain the same along $k_x$. Note that the discontinuity at 0º is due to slight photobleaching of the sample during the time course of the measurement.

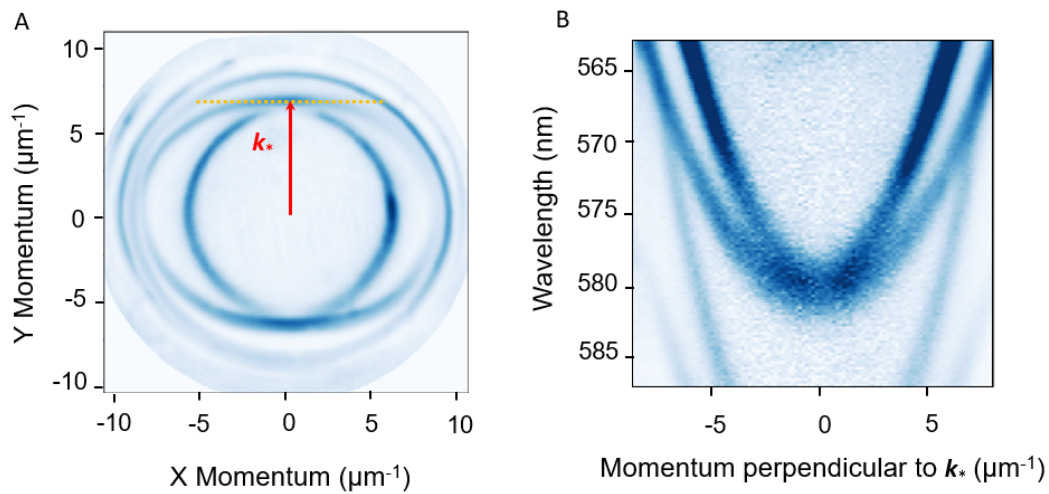

**Fig. S13. Details of polariton dispersions near the diabolic points for MAPbBr$_3$ in the orthorhombic phase.** (A) Constant energy cross section from an orthorhombic MAPbBr$_3$ microcavity showing the diabolical points. (B) The dispersion in the direction perpendicular to $k_*$ and intersecting the diabolical points shows two offset probolas.

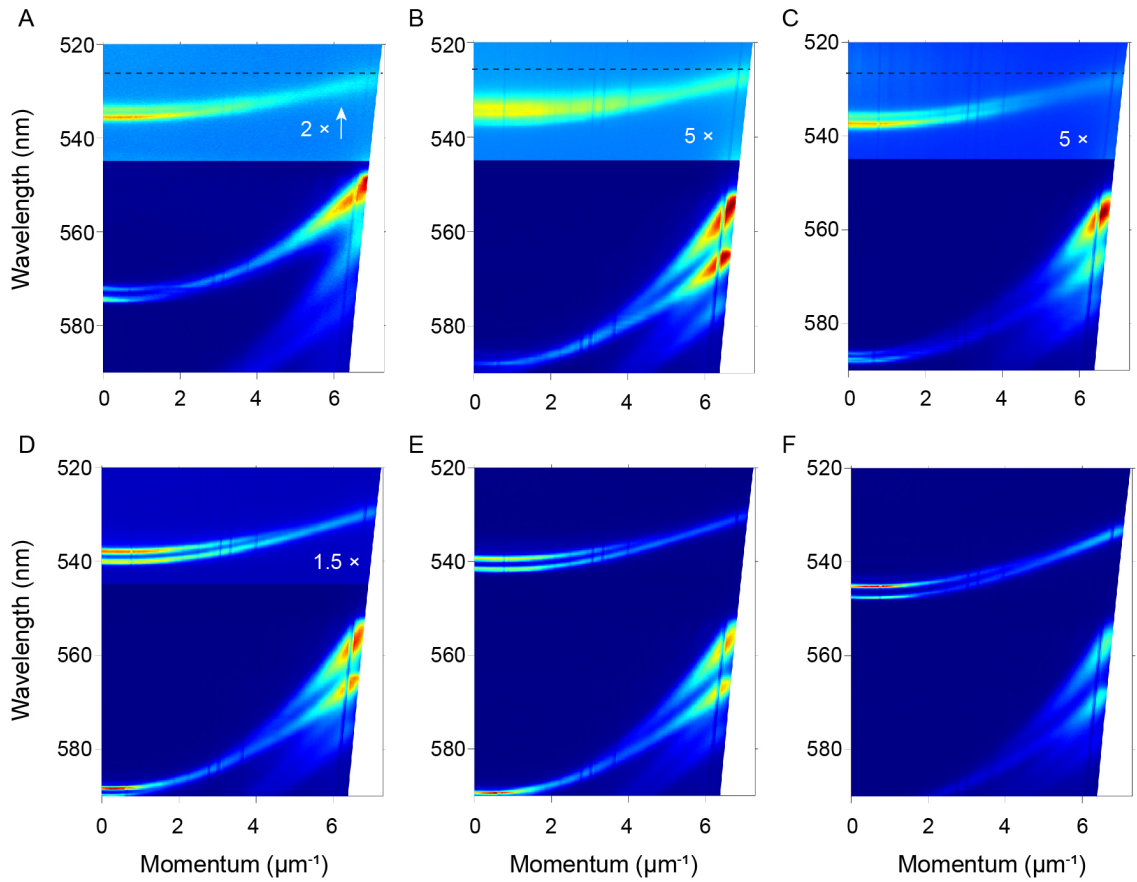

**Fig. S14. Dispersions collected on a series of crystals grown at different positions at room temperature**. These dispersions show deviations from parabolic dispersions at high *k* and avoided crossings with the exciton resonance for the higher energy polariton modes.

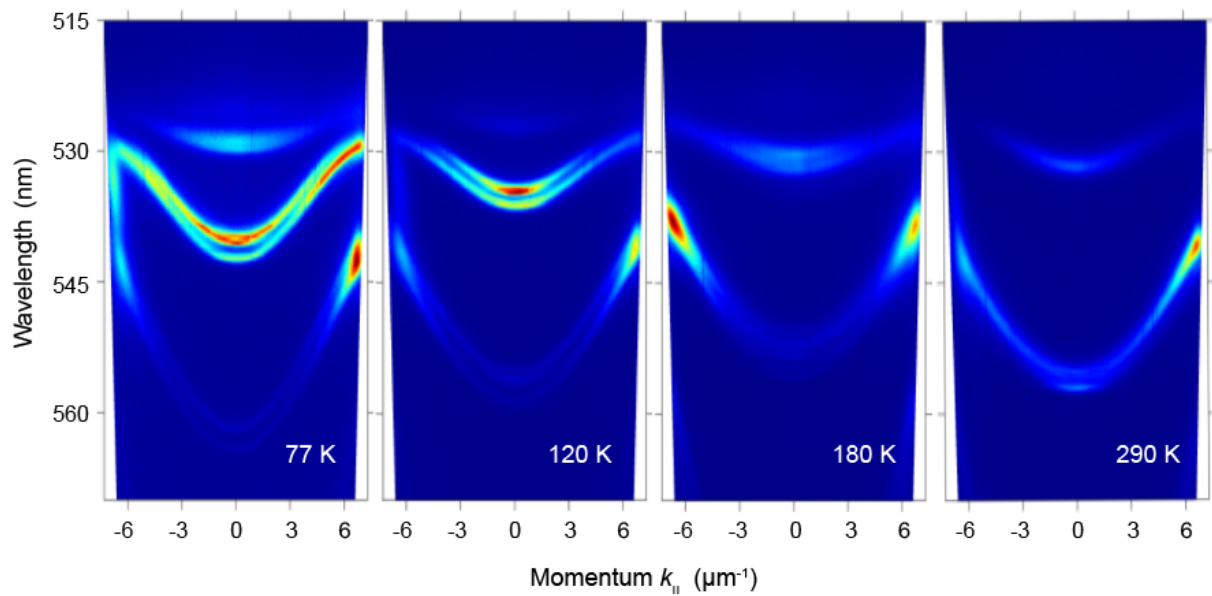

**Fig. S15. Temperature-dependent energy dispersion measured at a selected spot on CsPbBr$_3$ microcavity.** The evolution of the dispersion again confirms strong exciton-photon coupling presence at low temperature. It is also shown that that the strength of coupling gradually decreases as the temperature increases, due to the broadening of excitonic linewidth (as revealed by the temperature-dependent PL and absorption spectra in a previous study). Note that the energies of both cavity modes and excitons change when the temperature changes.

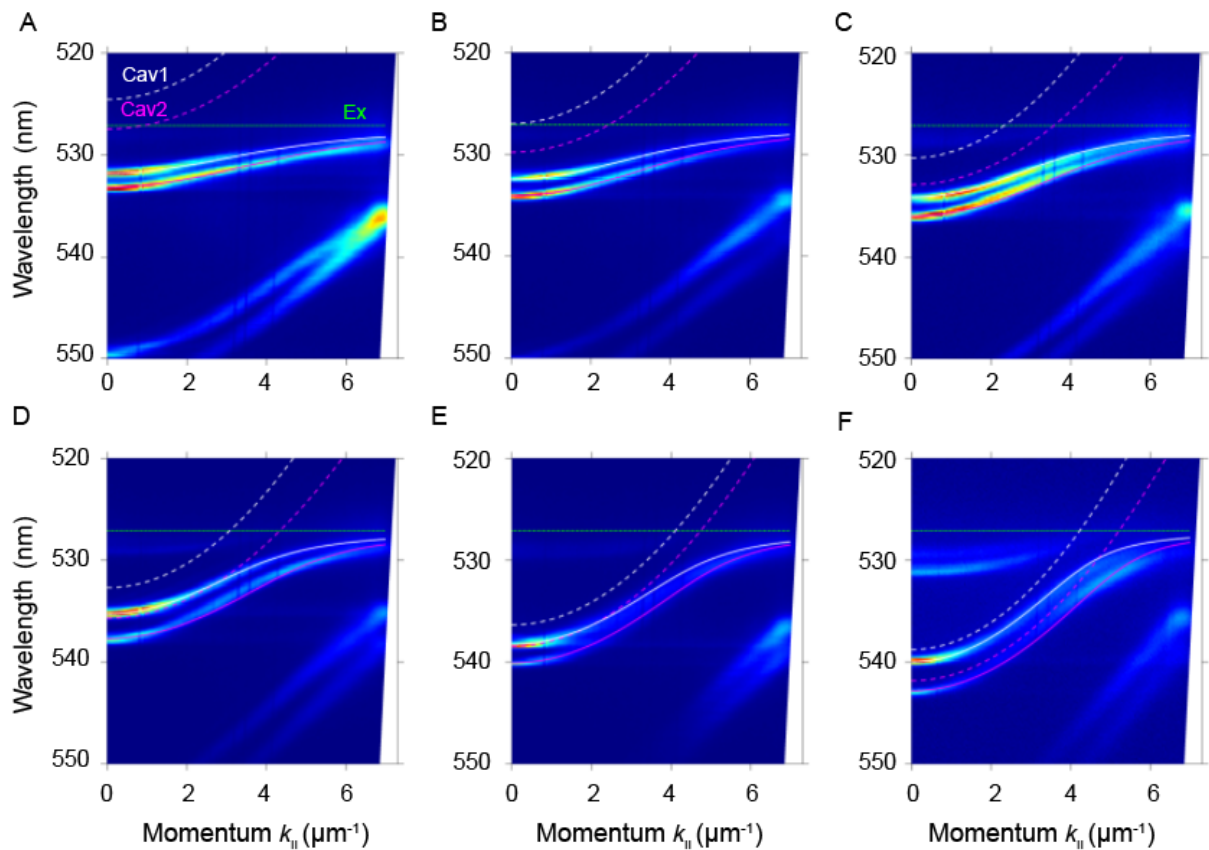

**Fig. S16. Dispersions collected on a series of crystals grown at different positions at 77 K, along with the corresponding models with the coupled oscillator, showing positive, resonant, and negative detunings for the lower polariton branches.** The white and magenta dashed lines are the dispersion of two SOC optical cavity modes. The solid lines are the modeled lower polariton branches using the coupled oscillator. The corresponding modeling gives $\Omega \sim 20\text{-}25$ meV.

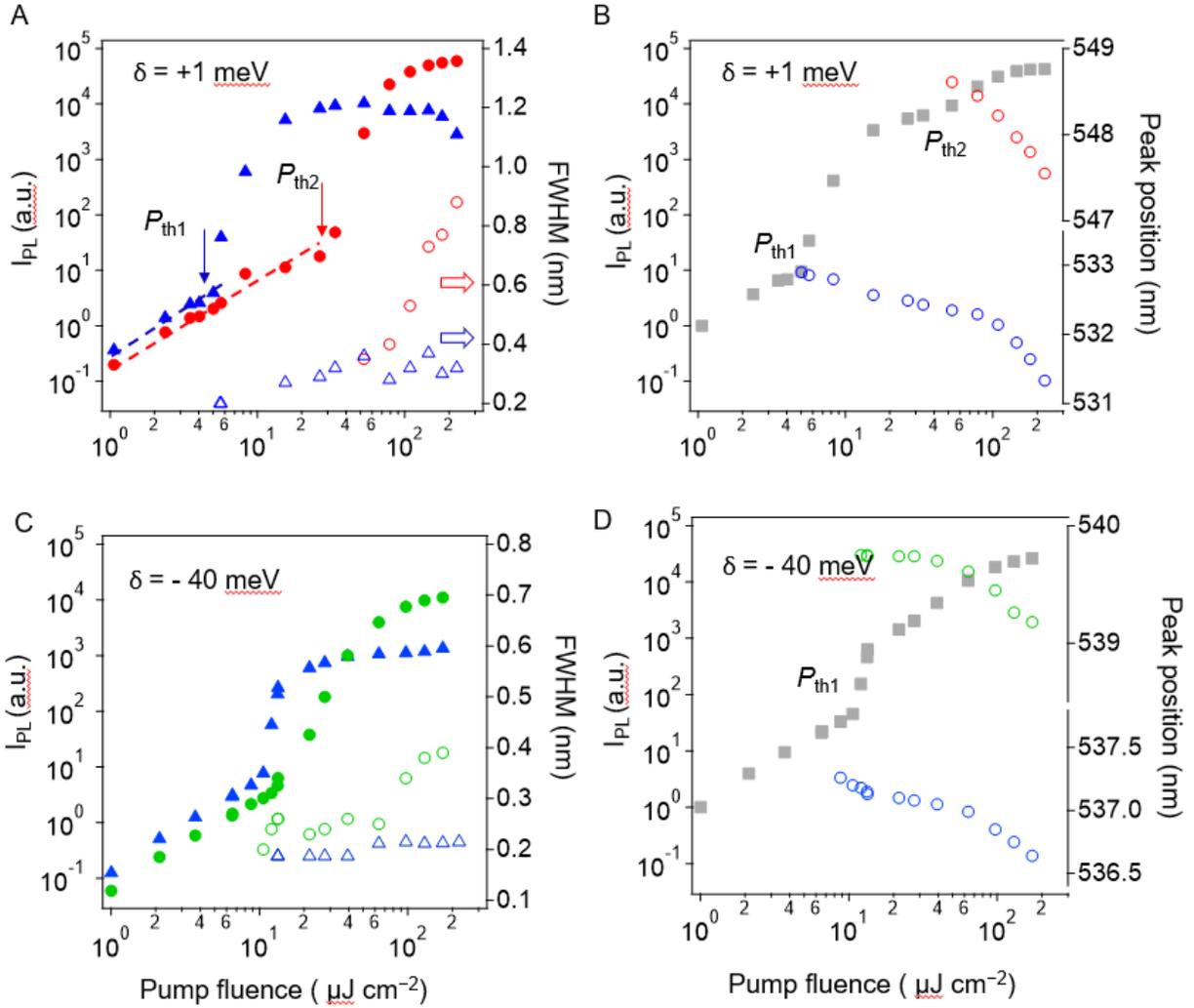

**Fig. S17. Pump power dependences showing the two lasing thresholds.** (A) Integrated PL intensity (left axis) as a function of $P$ in a log-log scale of the main lasing peaks (blue solid triangles, 530-534 nm; red solid circles, 544-551 nm), showing the two-threshold behavior for the $\delta = +1$ meV cavity. Also shown are the FWHMs (right axis) of the lasing peaks at ~533 nm (blue open triangles) and ~547 nm (red open circles). (B) Total PL intensity (525-555 nm) (grey squares, left axis) and lasing peak positions (open circles, right axis) as a function of $P$ for the $\delta = +1$ meV cavity. (C) $P$-dependences of integrated PL intensities (left axis) of the two lasing peaks from (blue solid triangles, 534-538 nm; green solid circles, 538-541 nm) and corresponding peak FWHMs (right axis; blue open triangles ~537 nm, green open circles, ~540 nm) for the $\delta = -40$ meV cavity. (D) Total PL intensity (525-555 nm) (grey squares, left axis) and lasing peak positions (open circles, right axis) as a function of $P$ for the $\delta = -40$ meV cavity.

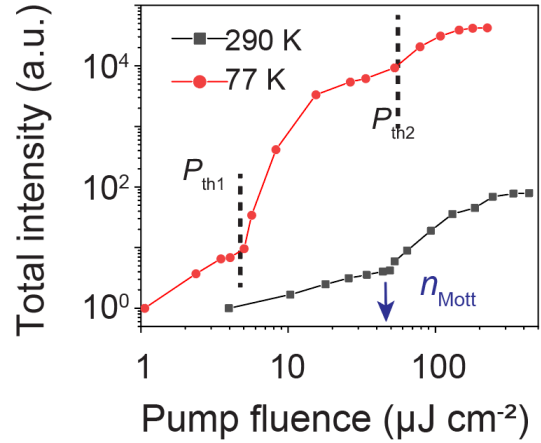

**Fig. S18. Integrated total PL intensity as a function of *P* in a log-log scale for a microcavity measured at room temperature (black curve), showing only one nonlinear threshold.** The low temperature at 77 K is also shown for comparison (red curve).

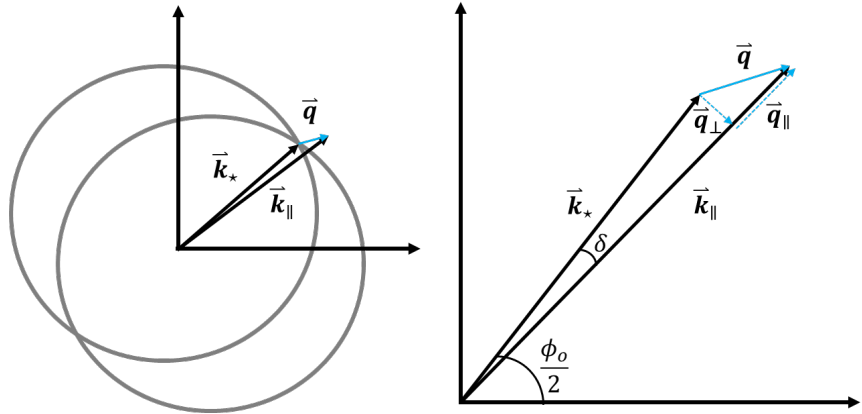

**Fig. S19. Demonstrating the expansion of the Hamiltonian about the Diabolical Point, in orders of a small vector q.** The left panel shows the two rings associated with the energy corresponding to the diabolical point, at an arbitrary crystal angle. The right panel shows the vectors associated with the expansion about this point. Note that the vector to the diabolical point makes an angle $\phi_o/2$ with the x-axis, and then the total wave vector makes the same angle with the x-axis, with a small deviation $\delta$. The right panel also demonstrates how, under the small angle approximation, the tangential component of the small wave vector q may be used to express the small angle.

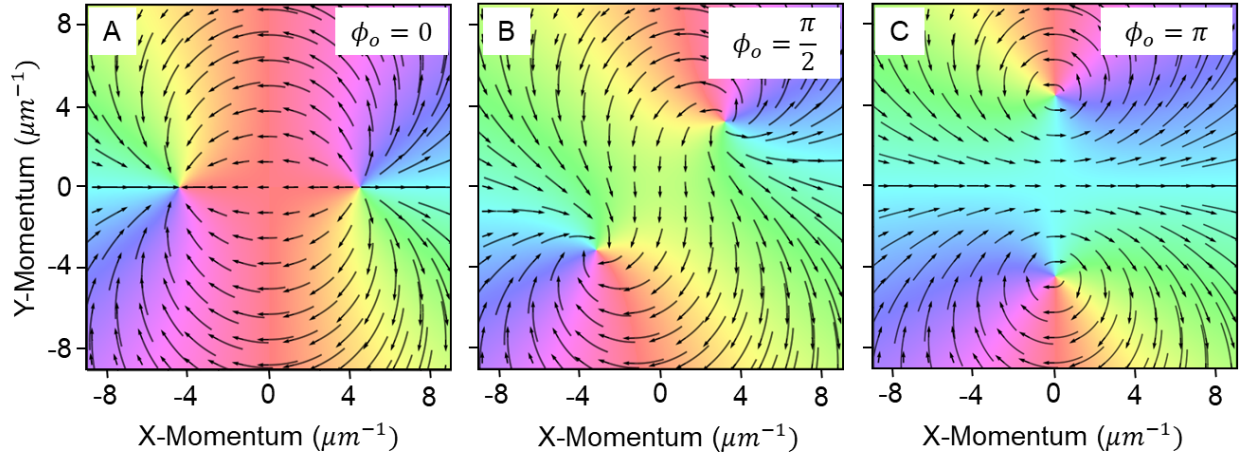

**Fig. S20. Dmonstrating the non-Abelian character of the effective magnetic field/ spin texture, and the role of crystal orientation on the spin texture about the diabolical point**. Panels A-C depict the field (vector field) as a function of the orientation of the anisotropy field, showing the phase of the lowest energy eigenvector (color) as a function of this orientation. We observe the divergence-like field in the case of the anisotropy field aligned along the x axis (A), and a curl-like field in the case aligned against the x axis (C). We observe an intermediate case in panel B, where the anisotropy field is aligned along the y axis.